 \documentclass[review]{cas-sc}

\usepackage[numbers]{natbib}

\usepackage{changepage} 
\usepackage{hyperref} %
\usepackage{tcolorbox}
\def\tsc#1{\csdef{#1}{\textsc{\lowercase{#1}}\xspace}}
\tsc{WGM}
\tsc{QE}
\tsc{EP}
\tsc{PMS}
\tsc{BEC}
\tsc{DE}

\begin{document}
\let\WriteBookmarks\relax
\def\floatpagepagefraction{1}
\def\textpagefraction{.001}

\shorttitle{Requirements Engineering for AI Systems}

\shortauthors{Ahmad et~al.}

\title [mode = title]{Requirements Engineering for Artificial Intelligence Systems:  \\ A Systematic Mapping Study}     



\author[1]{Khlood Ahmad}


\ead{ahmadkhl@deakin.edu.au}

\author[1]{Mohamed Abdelrazek}
\ead{mohamed.abdelrazek@deakin.edu.au}

\author[1]{Chetan Arora}
\ead{chetan.arora@deakin.edu.au}

\affiliation[1]{organization={Deakin University}, 
    addressline={ }, 
    city={Geelong},
    postcode={}, 
    state={VIC},
    country={Australia}}

\author[2]{Muneera Bano}
\ead{muneera.bano@csiro.au}

\affiliation[2]{organization={CSIRO's Data61}, 
    addressline={ }, 
    city={CLayton},
    postcode={}, 
    state={VIC},
    country={Australia}}

\author[3]{John Grundy}
\ead{john.grundy@monash.edu}

\affiliation[2]{organization={Monash University}, 
    addressline={}, 
    city={Clayton},
    postcode={}, 
    state={VIC},
    country={Australia}}

\begin{abstract}
~[Context] In traditional software systems, Requirements Engineering (RE) activities are well-established and researched.  However, building Artificial Intelligence (AI) based software with limited or no insight into the system's inner workings poses significant new challenges to RE.  Existing literature has focused on using AI to manage RE activities, with limited research on RE for AI (RE4AI).  [Objective] This paper investigates current approaches for specifying requirements for AI systems, identifies available frameworks, methodologies, tools, and techniques used to model requirements, and finds existing challenges and limitations.  [Method] We performed a systematic mapping study to find papers on current RE4AI approaches.  We identified 43 primary studies and analysed the existing methodologies, models, tools, and techniques used to specify and model requirements in real-world scenarios.  [Results] We found several challenges and limitations of existing RE4AI practices.  The findings highlighted that current RE applications were not adequately adaptable for building AI systems and emphasised the need to provide new techniques and tools to support RE4AI. [Conclusion] Our results showed that most of the empirical studies on RE4AI focused on autonomous, self-driving vehicles and managing data requirements, and areas such as ethics, trust, and explainability need further research.
\end{abstract}




\maketitle

\section{Introduction}\label{sec:introduction}

The increase in the volume and velocity of the data and the need for automation have made Artificial Intelligence (AI) a practical solution to many of the challenges we face today. This technology shift has allowed AI to become a favored software alternative to many organizations~\cite{holmquist2017intelligence}. Software systems with an AI component are currently under demand and are being investigated in multiple domains ranging from automotive and self-driving cars~\cite{schroeder2015design}, healthcare~\cite{jiang2017artificial}, science~\cite{li2017artificial}, virtual assistants chatbots~\cite{mekni2020smart, theosaksomo2019conversational}, predicting privacy decisions and providing privacy recommendations for IoT services~\cite{lee2019confident}. The process of building AI-based software differs from traditional software development approaches, making it more challenging to apply existing techniques and tools ~\cite{sculley2015hidden, lwakatare2020large}. Traditional software engineering involves collecting requirements, analysis, and detailed designs to implement an executable program which mainly includes writing code \cite{van2008software}. On the other hand, engineering software with an AI component involves additional configuration aspects, such as data collection, selecting an appropriate algorithm (e.g., a machine learning or natural language processing algorithm), and training the model based on the desired input/output with relatively less emphasis on the source code writing~\cite{sculley2015hidden, amershi2019software, jordan2015machine, arpteg2018software}.

The configuration process in AI-based software, i.e., selecting and validating data and other activities, need to be specified in the requirements phase. However, RE for AI systems (RE4AI) is not as established as the traditional RE approaches for non-AI software, and AI systems usually lack precise requirements and proper RE techniques~\cite{agarwal2014expert, khomh2018software}. The differences between the RE practices for traditional software (mostly deterministic) and AI-based software (mostly uncertainty-prone and black-box) create a need to adapt existing tools and RE methods~\cite{belani2019requirements}. 

AI has several definitions and encompasses several areas. John McCarthy, a pioneer of AI, defined AI in 1955 as ``the science and engineering of making intelligent machines''~\cite{ertel2018introduction}. The overall definition of AI is broad, and thus a number of sub-disciplines, such as Machine Learning (ML), Deep Learning (DL), Natural Language Processing (NLP), Expert Systems, and Robotics are considered under this broad umbrella term~\cite{ertel2018introduction}. In context of this paper, due to the lack of substantive research on RE for any individual sub-discipline, we do not focus on specific sub-disciplines and consider the AI systems as a whole.

In this article, we conducted a systematic mapping study to identify existing empirical evaluations, emerging theories, and occurring limitations and challenges in RE4AI. We observed that most of the results focused on using AI to manage RE~\cite{arora2015automated,abualhaija2020automated,ezzini2021using,zamani2021machine}, with little research supporting RE4AI. We also found most empirical studies focused on autonomous systems, with fewer empirical evaluations on ethics, explainability, and trust, and an apparent lack of research on managing data requirements and modeling requirements for AI systems.

We followed Kitchenham et al.'s~\cite{kitchenham2007guidelines} guidelines on performing systematic reviews and Petersen et al.'s~\cite{petersen2015guidelines} guidelines for conducting systematic mapping studies in software engineering to answer the following research questions~(RQs):

\begin{enumerate}[RQ1.]
    \item Which requirements engineering frameworks, notations, modeling languages, and tools have been proposed to build AI systems?
    \item Which evaluation methods are used to assess empirical studies in RE4AI?
    \item  Which target application domains and AI areas of focus are considered in the existing approaches?
    \item What are the limitations and challenges of existing requirements engineering techniques when applied to AI systems?
\end{enumerate}

We published preliminary results of our mapping study at the 29th IEEE Requirements Engineering (RE) conference~\cite{ahmad2021SLR}. The initial results were based on 27 studies published from 2010 to mid-2020. This article extends the results with an additional search between mid-2020 and mid-2021 to include 16 new research studies. One research question was added (RQ2), and R1, RQ3, and RQ4 were extended to include newer findings. RQ1 extends to include exiting frameworks and tools used to manage RE4AI, and the methods used to model requirements noticeably changed with the added studies.   In RQ3, we included which evaluation methods were more dominant in specific application domains. And finally, RQ4 results remained relatively similar, with additional papers to support existing issues and challenges and one new issue emerging. Also, this article provides a more comprehensive and detailed search method and lists two new research recommendations.        

The main research contributions of this mapping study include:
\begin{itemize}
\item  We identify a list of 43 primary studies that focus on RE for AI systems. Out of these 43 studies, 30 conduct some form of empirical investigation to either investigate a relevant problem or validate a solution. The remaining 13 studies are `non-empirical' papers that propose an idea or a model and do not conduct empirical investigation of any sort.
\item We report on \textit{three} existing frameworks, \textit{six} modeling languages or requirements notations, \textit{two} tools, and \textit{three} evaluation methods for RE4AI.
\item We identify the most popular domains that research RE techniques and domains that lack empirical research on RE4AI.
\item We identify \textit{nine} existing limitations and challenges in RE4AI.
\item We provide \textit{seven} recommendations for future research.
\end{itemize}

 The rest of the paper is structured as follows:  Section~\ref{sec:Background}
provides a brief background on RE4AI. Section~\ref{sec:SearchMethod} presents details of our structured literature search and selection process on RE for AI systems. Section~\ref{sec:Results} reports results from the selected primary studies. Section~\ref{sec:Threats} addresses threats to validity. 
 Section~\ref{sec:Discussion} discusses key results and summarizes emerging theories, and Section~\ref{sec:Conclusion} concludes.

\section{Background and Related Work}~\label{sec:Background}

\subsection{Requirements Engineering}

RE is arguably the most crucial phase in the software engineering lifecycle and plays a significant role in every stage of software development~\cite{dick2017requirements}.  In RE, understanding stakeholders' requests are important, and requirements act as the communication channel between the system developers and the stakeholders \cite{wheatcraft2018communicating}.  RE acts as that channel to gather and document stakeholders' needs~\cite{nuseibeh2000requirements}.  Therefore, it is vital to establish requirements early on when building software systems to ensure all stakeholders' needs and specifications are captured and documented correctly.

Koelsch defined requirements as ``a need, desire, or want to be satisfied by a product or service"~\cite{koelsch2016requirements}.  If this need or desire is not fulfilled, the product is not usable.  An established process for RE is obtained to achieve this need and undergoes the following phases:  elicitation, analysis, specification and documentation, validation, and management~\cite{inayat2015systematic}.  This process makes sure that requirements are extracted, documented, managed correctly, and comply with users' needs.  However, the software process differs when AI components are involved.  Specifications do not always drive such systems, as parts of the system can be driven from data~\cite{hu2020towards}.  Therefore, RE techniques would need to be adjusted to the changes introduced by this new paradigm of AI systems.

In RE, requirements are classified to be either functional or non-functional.  Functional requirements represent the system's features and business rules to what the system should include.  Whereas non-functional requirements include systems qualities and constraints~\cite{koelsch2016requirements}.  Non-functional requirements (NFR) are more difficult to present~\cite{lapouchnian2005goal}.  However, several modeling languages and tools are available that can help display properties of non-functional requirements. 

Modeling languages display WHY behaviors and functionalities are selected and WHAT capabilities are needed to support these choices.  Modeling languages focus on the high-level abstraction aspect of the required system rather than the details of operations, which is helpful during the early stages of building software systems~\cite{amyot2011user, gonccalves2019understanding}.  There are different RE modeling languages available such as Goal-Oriented Requirements Engineering (GORE) and User Requirements Notation (URN)\cite{amyot2011user}. Goal modeling languages can help present requirements that showcase the stakeholders objectives of what is needed from the proposed system \cite{duran2019reusability}.

\subsection{Requirements Engineering and AI-based Systems}

When engineering software systems with an AI component, new processes are appearing, such as managing data, training the models and the design process \cite{kuwajima2020engineering, bosch2021engineering}.  In SE, the ML code is relatively small compared to the actual process when building systems with ML components.  Most of the work focuses on managing data, feature extraction, analyzing, configuring, etc.~\cite{sculley2015hidden}.   At Microsoft, the SE process is employed over a nine-step model.  It includes: gathering requirements, collecting, cleaning, and labeling data, feature engineering, training and evaluating the model, and finally deploying and monitoring the model over time.  Requirements are decided based on how feasible it is to implement features and find appropriate models for given problems
\cite{amershi2019software}. 

Because of the difference in the development process, new challenges are appearing when managing requirements for AI software, some of these challenges include defining, eliciting, and specifying requirements. {Kuwajima et al.~\cite{kuwajima2020engineering} explains that the lack of requirements specifications in current Machine Learning (ML) systems significantly impacts the ML model's quality. And that most ML models lack requirements specifications. One of the reasons for the difficulties in writing requirements specifications for AI software is the inconsistencies in inputs and outputs patterns. Several tools are used for traditional SE practices to manage code and other issues. However, because of the vast difference between AI software and traditional SE processes, it is hard to use these tools in managing such issues~\cite{sculley2015hidden}.

From the literature, it is evident that the existing tools for traditional SE practices (more specifically RE) cannot be appropriately utilized when building AI systems~\cite{sculley2015hidden}. Also, including AI components in building software systems has impacted RE, and new requirements have appeared in the process, such as data, ethics, explainability, and trust.  Some existing requirements have changed. For example, Non-functional requirements (NFR) for ML systems have changed to include transparency, trust, privacy, safety, reliability, and security~\cite{cysneiros2018software}. From a RE perspective, we are facing a new set of challenges and questions when building AI systems~\cite{houdek2017automotive}.  Specifying and defining requirements has introduced new challenges in RE practices~\cite{lwakatare2019taxonomy,  kuwajima2020engineering}.  This motivated us to analyze the existing body of work and collate the state of the art of empirical literature on RE4AI.

\subsection{Related Work}

RE4AI has gained traction from the research community in the past few years. 
Previous studies (on RE4AI or Software Engineering for AI (SE4AI) with RE recommendations) focused on showing the difference between traditional software and AI software~\cite{kostova2020interplay}, identifying issues and challenges in RE4AI, and reporting on research directions for future development~\cite{chazette2019mitigating, kaindl2020towards, heyn2021requirement, camilli2021risk, odong2022requirements}. Other studies investigated the use of non-functional requirements and issues related to non-functional requirements~\cite{cysneiros2018software, brugali2019non}, ethical requirements \cite{guizzardi2020ethical, crnkovic2012robots, thinyane2020multi}, explainability \cite{nguyen2021holistic}, and transparency~\cite{felzmann2019transparency}. A number of authors have investigated what methods and tools could be used in RE4AI. In \cite{agarwal2014expert}, the study suggests a list of techniques that can be used for each RE phase when working with expert systems, and \cite{fagbola2019towards} lists a number of tools that can identify and mitigate any issues related to AI fairness and bias during RE. 

We have further identified two relevant mapping studies. The first study identified 348 existing SE4AI studies and provided some insights on RE-related issues~\cite{martinez2021software}. This study only presents a general overview and does not provide details on the current methods, frameworks, notations, or tools used to manage RE4AI. The second mapping study, conducted by Villamizar et al.~\cite{villamizar2021requirements}, identified 35 RE for ML-related studies and provided an overview of some of the existing challenges, popular RE phases, and existing evaluations. The findings showed that 40\% of the selected papers did not have any form of empirical investigation, similar to our results (discussed later). Our study complements the second mapping study, 
and sheds light on additional aspects of RE4AI (in addition to the second study), such as identifying existing methodologies, tools, and modeling notations used.

\section{Systematic Mapping Study}~\label{sec:SearchMethod}

This section shows our search strategy to extract relevant studies based on Kitchenham et al.'s~\cite{kitchenham2007guidelines} guidelines to conduct an SLR and Petersen et al.'s  \cite{petersen2015guidelines} guidelines for conducting systematic mapping studies in software engineering.  The mapping study followed three stages to include planning, conducting, and reporting the review.  The initial planning phase involved writing a protocol and identifying a set of research questions.  The protocol\footnote{\href{https://www.dropbox.com/s/mp6r9l4rpao4hsr/SLR\%20Protocol\%20Khlood.pdf?dl=0}{Protocol Link}} included a plan for a search strategy.   To exhaust our exploration of any existing empirical evidence, we identified relevant keywords and search strings.  The second stage involved identifying and analyzing existing primary studies to answer our research questions. The final step involved evaluating, reviewing and reporting the final document.  We performed the search over two periods.  The first period was from 2010 and mid-2020, and the second was from mid-2020 to mid-2021.  We then combined the papers resulting from both searches as displayed in Figure.~\ref{fig:SearchProcess}. 

\begin{figure}[h]
\centering
\includegraphics[width=\linewidth]{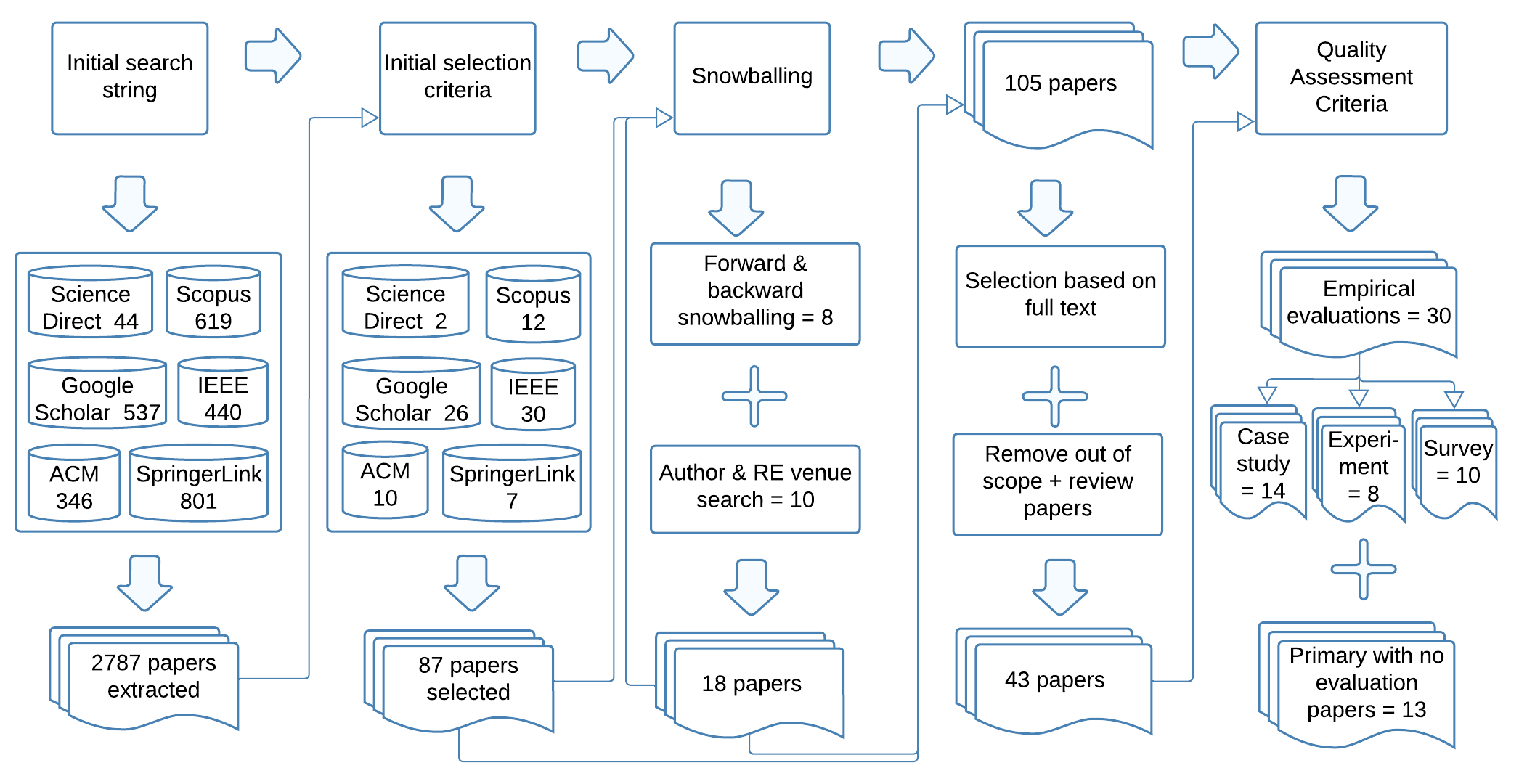}
\caption{Paper extraction process}
\label{fig:SearchProcess}
\end{figure} 

\subsection{Search Strategy}

Our research questions guided the identification of the main search terms to include ``Requirements Engineering'' and ``Artificial Intelligence''.  We extracted synonyms and alternative words from each search term and customized a search string for each database. For both main search terms,  we derived several alternative words as shown in Table~\ref{table:searchTerms}.  For ``requirements engineering'', the search terms were derived from existing research on RE \cite{bano2014systematic, ambreen2018empirical} and meetings conducted between the first and fourth authors to finalize the list of keywords used.  Keywords selected for ``Artificial Intelligence'' were based on a combination of existing literature \cite{hager2017advances, stone2016artificial} and meetings conducted between the first and second authors to decide on which keywords to include. 

\begin{table}[h]
\caption{List of Search Terms} 
\label{table:searchTerms}
\centering
\footnotesize
\begin{tabular}{p{1.5cm} p{12cm}}
\hline
\textbf{Main Terms } & \textbf{Alternative terms}  \\
\hline     
requirements engineering        &   requirements process, requirements elicitation, requirements gathering, requirements identification, requirements analysis, requirements validation, requirements verification, requirements specification, requirements development,  requirements documentation,  requirements management, requirements testing, requirements driven, functional requirements   \\
\hline     
artificial    intelligence &  machine learning, speech recognition, deep learning, natural language processing, computer vision, machine intelligence, chatbot, chat-bot, expert systems, self driving,  autonomous,  recommendation system,  robot, AI,  ML   \\
\hline
\end{tabular}
\end{table}

We used boolean operators to link the key terms and their synonyms when connecting the search string.  As a result, we formed the following search string:
\vspace*{0.75em}
\begin{adjustwidth}{1cm}{1cm}
ON ABSTRACT ((``requirements engineering'' OR ``requirements process'' OR ``requirements elicitation'' OR ``requirements gathering'' OR ``requirements identification'' OR ``requirements analysis'' OR ``requirements validation'' OR ``requirements verification'' OR ``requirements specification'' OR ``requirements development'' OR ``requirements documentation'' OR ``requirements management'' OR ``requirements testing'' OR ``functional requirements'' OR ``requirements driven'') AND (``artificial intelligence''  OR ``machine learning'' OR ``expert systems''  OR ``deep learning''  OR ``computer vision'' OR ``natural language processing'' OR ``speech recognition'' OR ``machine intelligence'' OR AI OR ML OR chatbot  OR ``expert systems'' OR ``self driving'' OR autonomous OR ``recommendation system'' OR robot)). \\
\end{adjustwidth}

We searched the following online databases: IEEE Explore, ACM Digital Library, Google Scholar, Science Direct, SpringerLink, and Scopus.     For most databases, the search string was conducted based on abstracts.  However, some databases allowed us to include titles and keywords, and others, such as Google Scholar, limited search results on abstract to the last year only.  For Google Scholar, the initial search returned 17000 results.  To narrow them down, we first performed a search based on abstracts, and to cover as many results as possible, we then completed another search on titles only.  When entering the search string, we found that most online databases had constraints and limits.  Most searches limited either the number of words used, the number of boolean connectors or the length of the string itself.  Therefore, the search string was modified to accommodate each database's search criteria while keeping the logical order consistent.  Initial tests on the search strings were carried out to check the resulting papers, and the results were verified by the second and fourth authors. When testing the search string we found that some of the papers that we had extracted prior to the search did not appear in our search. Thus, we kept modifying the search string and keywords until all known papers were found. For example, we could only find the paper \cite{bosch2018takes} after including the keywords ``requirements driven'' to the search string.

\subsection{Selection Criteria and Snowballing}

The initial search resulted in 2,787 papers.  To filter the results further, we performed an initial selection criteria to extract relevant studies.  We assessed each article based on its title and abstract and performed an inclusion and exclusion assessment as shown in Table~\ref{table:inclusionExclusion}.  A folder was created for each database, and papers that passed the inclusion criteria were given an identifier number and placed in the corresponding folder.  A total of 26 papers were found in more than one database so we removed duplicates from the final list. For example, ~\cite{cysneiros2018software} was found in IEEE, Google Scholar, and Scopus, thus, we removed the duplicates found in Google Scholar and Scopus.  For studies that reported several publications,  we grouped them under the same identifier.  After completing the initial selection criteria, we found a total of 87 studies. 

\begin{table}[htbp]
\caption{Initial inclusion and exclusion assessment criteria } 
\label{table:inclusionExclusion}
\footnotesize
\begin{tabular}{p{6cm} p{7cm}}
\hline
\textbf{Inclusion criteria:}  & \textbf{Exclusion criteria:} \\ \hline
Papers published in English language &  Review papers and secondary studies \\ \hline
Primary studies on requirements engineering for AI systems & Studies that used AI to manage or analyze requirements \\ \hline
Papers published at peer reviewed conferences and journals & Abstract-only papers -- the full paper is not available or only found as an arXiv pre-print (Not published) \\ \hline
Full resource papers available & Exclude book chapters, magazines, general articles, project plans or theses \\ \hline
\end{tabular}
\end{table}  

Next, we performed backward snowballing on all the selected papers by examining the references and forward snowballing by checking the citations and authors \cite{wohlin2014guidelines}.  The backward snowballing involved extracting relevant research from references.  Whereas, the forward snowballing reviewed all the papers that cited the selected studies. We examined all the references and citations of the 87 articles resulting from the initial selection criteria and searched the author's Google Scholar profiles and the proceedings of RE-related publication venues. A total of 18 new papers were found during snowballing. The same backward and forward snowballing method was applied to the identified 18 papers.  However, no new papers were identified in the process.

Next, the first author read the full text on all 105 papers to determine if they were relevant to our RQs.  This selection process was carried out in several meetings among the first four authors to decide on which papers to include.  Furthermore, papers that felt out of context were flagged by the first author and validated by the second and third authors. Out of these papers, we excluded 26 secondary papers, i.e., papers that report their findings based on existing primary studies, such as literature reviews. From the remaining studies we removed 23 of the papers as they were not relevant, did not answer our RQs, or were out of scope. We further excluded one paper, that on closer analysis focused on AI for RE, instead of RE4AI. Nine more papers were excluded because the focus of the proposed method and evaluation was on non-AI related tasks. For example, in \cite{lee2019confident}, the paper is excluded as the proposed requirements are for the Internet of Things (IoT) aspect of the project and not the ML model used. Also, we excluded two arXiv papers that were not yet published. However, we did include papers that appeared as an arXiv pre-print but were peer-reviewed and accepted at a later dated conference. The remaining 44 papers were identified as primary studies. Two of these papers reported the same study under two different publications so we only included the most recent publication. Thus, a total of 43 papers were selected for quality assessment.

\subsection{Quality Assessment} 

Of the total 43 papers, we classified them as `empirical' or `non-empirical' papers as shown in Table \ref{table:classifications}.  We found that 13 of the primary studies did not have an evaluation component, i.e., they proposed an idea, model or a solution, but did not present any evaluation of the proposed artefact or left the evaluation for future work.  We classified these papers as `non-empirical' and used the quality assessment check in Table~\ref{table:QANonEmpirical}.  The remaining 30 papers that had an evaluation component, e.g., conducted an experiment, survey or a case study, were classified as `empirical', and we performed the quality assessment check based on Kitchenham’s guidelines as shown in Table~\ref{table:QAC}. We note that we did not discard any paper in quality assessment (due to the already limited number of studies on the topic), and the assessment helped in ranking the primary studies.

\begin{table}[h!]
\caption{Classifications used for selected primary studies} 
\label{table:classifications}
\footnotesize
\begin{tabular}{p{2cm} p{5cm} p{7cm}}
\hline
\textbf{Classification} & \textbf{Explanation}& \textbf{Example}\\ \hline
Non-empirical & Idea paper with no evaluation & K\"{o}hl et al. \cite{kohl2019explainability} present a conceptual analysis for explainability as a NFR and proposed to model them using “Softgoal Independency Graph SIG” with other NFR and minimize conflicts \\
 & Propose a new model, or prototype with no evaluation & Aydemir and Dalpiaz \cite{aydemir2018roadmap} provide an “Ethics-Aware SE Method” but is not evaluated\\ \hline
Empirical & The evaluation component is used to validate the proposed RE4AI method, idea or tool & Nalchigar et al. provide a new ``GR4ML” framework that is evaluated using a case study \cite{nalchigar2021modeling} \\
 & The evaluation is used to propose a model or investigate an RE4AI problem, & Vogelsang and Borg \cite{vogelsang2019requirements} interviewed practitioners to investigate the challenges in RE for ML systems \\ \hline
\end{tabular}
\end{table}

The different types of empirical investigations~\cite{wohlin2012experimentation}, found in 30 `empirical' studies are: 

\begin{enumerate}
    \item \textbf{Surveys}: These included collecting data from a selected population sample using methods such as questionnaires and interviews, and the main purpose of a survey was to answer questions to a given problem \cite{glasow2005fundamentals, gable1994integrating}.  We checked if the sample size was justified, inclusive and if any biases were evident in the selection process. 
    \item \textbf{Case studies}: Case studies included in-depth investigations to test a new or existing theory based on a set of research questions that aim to address the \emph{how} and \emph{why} a phenomenon works \cite{tellis1997application, meyer2001case}.  Data collection methods in case studies included surveys and interviews with participants or using data from existing repositories \cite{wohlin2021case}.  We noted if the process had a clear chain of evidence and if the context was clearly defined.  We also checked for any ethical issues related to participants or data selection.  Are the participants adequately described, and if the data is adequately presented.
    \item \textbf{Experiments}: Selected experiments involved testing a hypothesis with one or more independent variables against dependant variables.  For example, testing out a newly proposed method against the traditional approach to carrying out a procedure \cite{easterbrook2008selecting}.  We checked if the sample size was justified, randomly selected, without biases and if they answered the research questions. 
\end{enumerate}



\begin{table}[htbp]
\caption{Quality Assessment Criteria for Empirical Papers based on \cite{kitchenham2007guidelines}. Each question is given a grade of either Yes, No or Partly (Y/N/P)} 
\label{table:QAC}
\footnotesize
\begin{tabular}{p{2cm} p{9cm} p{1cm} }
\hline
\textbf{Study type} & \textbf{Question} & \textbf{Grade} \\ \hline
General & Are the aims clearly stated? & Y/N \\  & Are the measures used clearly defined? & Y/N/P  \\  & Is the methodology clearly described? & Y/N/P  \\    & Data collection methods adequately described? & Y/N/P  \\  & Does the report have any implication for practice? & Y/N/P  \\  & Do the researchers explain the consequences? & Y/N/P  \\ \hline
Survey & Was the sample size justified?& Y/N  \\  & Is the sample representative of the population? & Y/N/P  \\  
& Is the survey likely to have introduced bias?& Y/N/P  \\  
& Is there a comparison or control group?& Y/N/P  \\  & Evidence of selection biases in the group selected?   & Y/N/P  \\ \hline
Experiments & Was the sample size justified?  & Y/N         \\   & Are the research questions answered? & Y/N/P  \\  & Were the experiments randomly allocated?& Y/N         \\ 
& Choice of subjects influence the outcome?& Y/N/P  \\ 
& Could lack of blinding introduce bias?& Y/N/P  \\ \hline
Case study & Is the context of the case study defined? & Y/N         \\  
 & Are the study participants adequately described?  & Y/N/P  \\ 
 & Are sufficient raw data presented? & Y/N/P  \\  & Are ethical issues addressed properly? & Y/N/P  \\ 
 & Is a clear Chain of evidence established from observations to conclusions? & Y/N/P  \\ \hline
 
\end{tabular}
\end{table}

\begin{table}[]
\caption{Quality Assessment Criteria for Non-Empirical papers} 
\label{table:QANonEmpirical}

\begin{tabular}{p{7cm} p{1cm}}
\hline
\textbf{Question} & \textbf{Grade} \\ \hline
Does the study propose an idea or a model? & Yes/No \\ 
Are the aims clearly stated? & Yes/No \\
Are the measures used clearly defined? & Yes/No \\ 
Does the report have any implication for practice? & Yes/No \\ \hline
\end{tabular}
\end{table}

For empirical evaluation, checklist items were graded by yes, no, or partial.  Each item had a score of 1 for yes, 0 for no, and 0.5 for partial.  After finalizing the quality assessment criteria, we selected any study that scored more than~4.  The final score for empirical evaluations was out of 11.  We labeled papers with a score between 4 and 6 as low.  Medium for scores between 6.5 and 8.5.  And high for scores between 9 and 11 as shown in Figure.~\ref{fig:QualityAssessmentScore}.   For papers that did not have an evaluation method, the quality was assessed based on the criteria shown in Table~\ref{table:QANonEmpirical}. We kept the papers that scored low on the quality assessment due to the limited available work.  Also, we felt that each of these papers could contribute to our findings. A total of 30 empirical studies and 13 non-empirical studies were selected as shown in Table~\ref{table:selectedPapers}.

\begin{table}[h!]
\caption{Selected primary studies} 
\label{table:selectedPapers}

\begin{tabular}{p{2cm} p{9cm}}
\hline
\textbf{Classification} & \textbf{Selected Papers} \\ \hline
Empirical & \cite{bonfe2012towards, hu2020towards, gruber2017integrated, sandkuhl2019putting, rahimi2019toward, jakob2017defining, vogelsang2019requirements, tuncali2019requirements, becker2020partial, bosch2018takes, shin2019data, dimitrakopoulos2019alpha, weihrauch2018conceptual, fenn2016addressing, neace2018goal, lockerbie2020using, nakamichi2020requirements, challa2020faulty, ries2021mde, olmos2020helping, samin2021towards, cirqueira2020scenario, ntakolia2020user, nalchigar2021modeling, schoonderwoerd2021human, islam2021mobile, cleland2020requirements, hall2019systematic, habibullah2021non, rivero2021lessons}\\ 
Non-empirical  & \cite{horkoff2019nonFunctional, aydemir2018roadmap, bruno2013functional, kohl2019explainability, altarturi2017requirement, ang2011requirement, ishikawa2020evidence, kuwajima2019adapting, amaral2020ontology, agrawal2020model, clauer2021usage, khatamino2021nlp, schwammberger2021quest} \\
\hline
\end{tabular}
\end{table}

\begin{figure}[htbp]
   \centering
\includegraphics[width=0.8\linewidth]{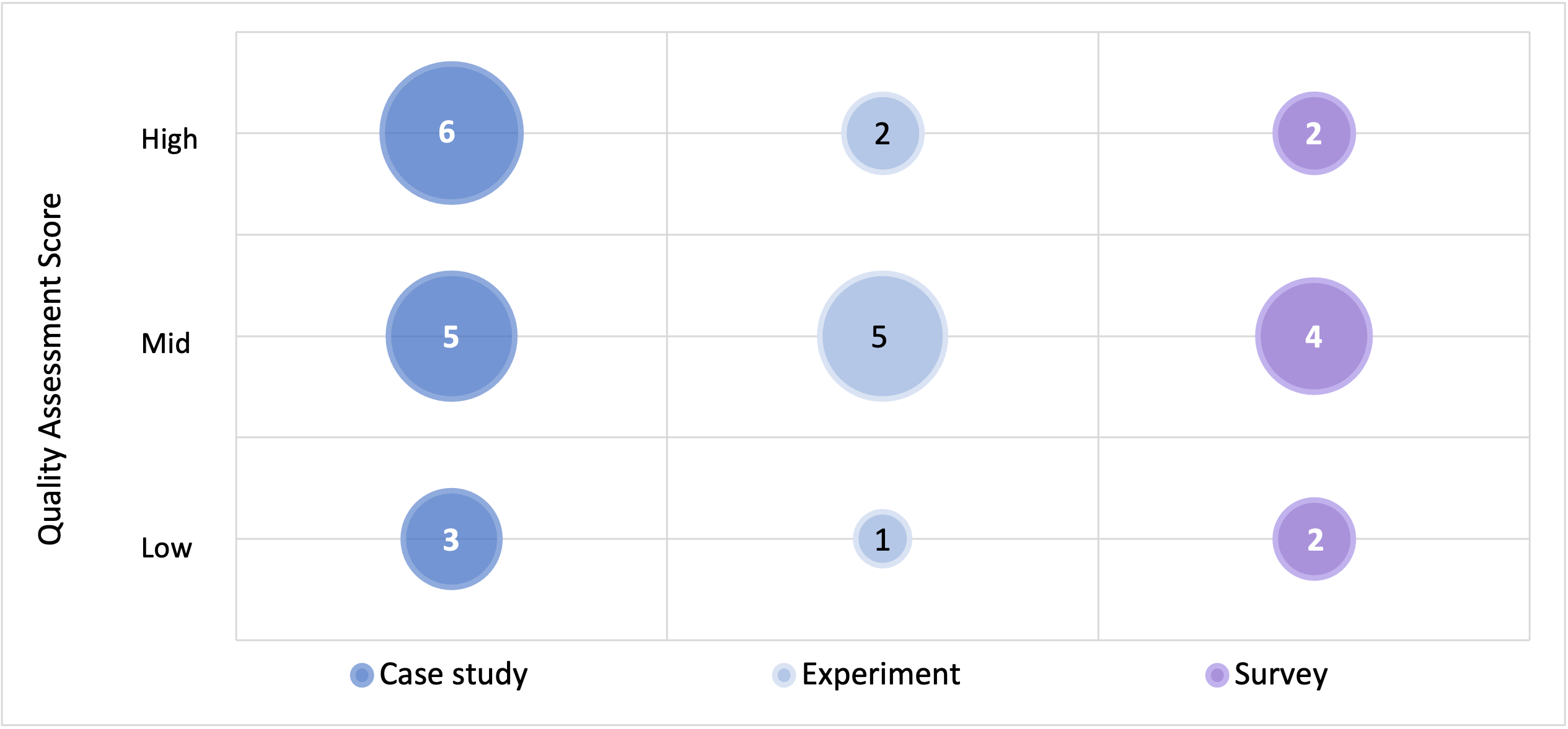}
\caption{Distribution of the resulting empirical papers after performing the quality assessment score}
\label{fig:QualityAssessmentScore}
\end{figure}

\begin{figure}[htbp]
\centering
\includegraphics[width=\linewidth]{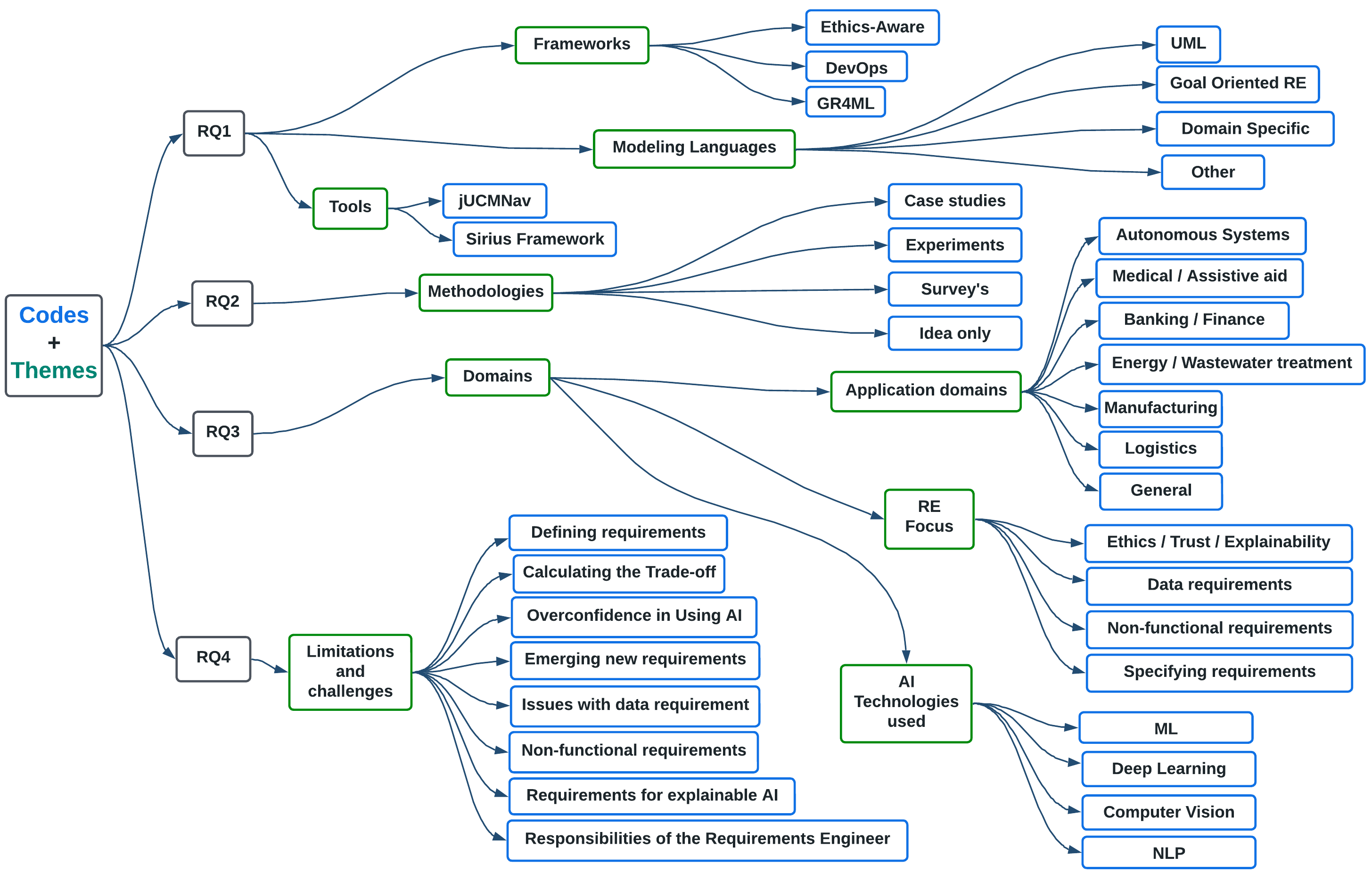}
\caption{Codes and themes emerging during data extraction based on our research question's}
\label{fig:CodeGeneration}
\end{figure}

\subsection{Data Collection and Analysis } 

We had several meetings between the first, second, and fourth authors to decide on what information to extract from each of the selected studies. A shared Excel document\footnote{\href{https://www.dropbox.com/s/wczhl5b5likyfv6/SelectedPapers.xlsx?dl=0}{Data extraction for selected papers}} was created, and each paper was given a unique ID. We extracted publication details that included:  title, authors' names, citation count, source, type (conference, workshop, or journal), and ranking. We then documented the type of the study, the proposed framework (if any), and the evaluation methodology, e.g., case study, experiment, survey, or a `non-empirical' paper. We further extracted the data collection methods used in empirical studies and the modeling tools or requirements notations used in each paper. We further noted the application domain (e.g., medical and autonomous systems). Next, we listed the publication date for each paper to determine if there was a trend in the number of publications per year on RE4AI related research.

For data analysis, we conducted a `theoretical' thematic analysis that was driven by the research questions (RQs)~\cite{braun2006using}. The codes were initially selected based on the RQs in a couple of meetings between the first and the second authors. Some of the codes emerged during the coding process. For example, we did not know what limitations or challenges existed in the literature, so the codes and themes were established as they appeared.  
 Once the codes were established, we entered the selected 43 papers into Nvivo for coding. The main codes and sub-codes were created in Nvivo, and the first author read through each paper and used an open-coding procedure on each transcript to assign relevant text to its matching code, and the data was extracted with a top-down coding strategy~\cite{urquhart2012grounded}.   The third author closely reviewed the coding results for a randomly selected sample of $\approx$25\% papers. They expressed only minor disagreement on the coding results for one paper in terms of limitations and challenges. We discussed this, and however, no changes were made in the coding results after agreement. Once the coding process was completed, we combined relevant codes into themes, as shown in Figure.~\ref{fig:CodeGeneration}. The final themes were then presented and discussed in a number of meetings among the first, second, and third authors for review and analysis. We then used these themes to answer our RQs and present any emerging theories.

\section{Results}~\label{sec:Results}

During the initial search, we found that more studies focused on using AI to manage RE and less work on RE4AI.  For example, from the first ten results returning from the IEEE Xplore database during the initial search, eight of these papers researched ways to manage RE using AI, and only two focused on RE4AI.  Similar patterns were evident in most of the search results obtained from other databases.  We also observed that the amount of research in RE4AI has increased lately, as shown in Figure.~\ref{fig:PPY}.  We found that 90\% of the primary studies were published between 2017 and 2021, with 74\% of the results published in the last three years.  The increase in publications indicates that more researchers are looking into addressing RE-related issues when building AI systems. 

\begin{figure}[htbp]
   \centering
\includegraphics[width=0.7\linewidth]{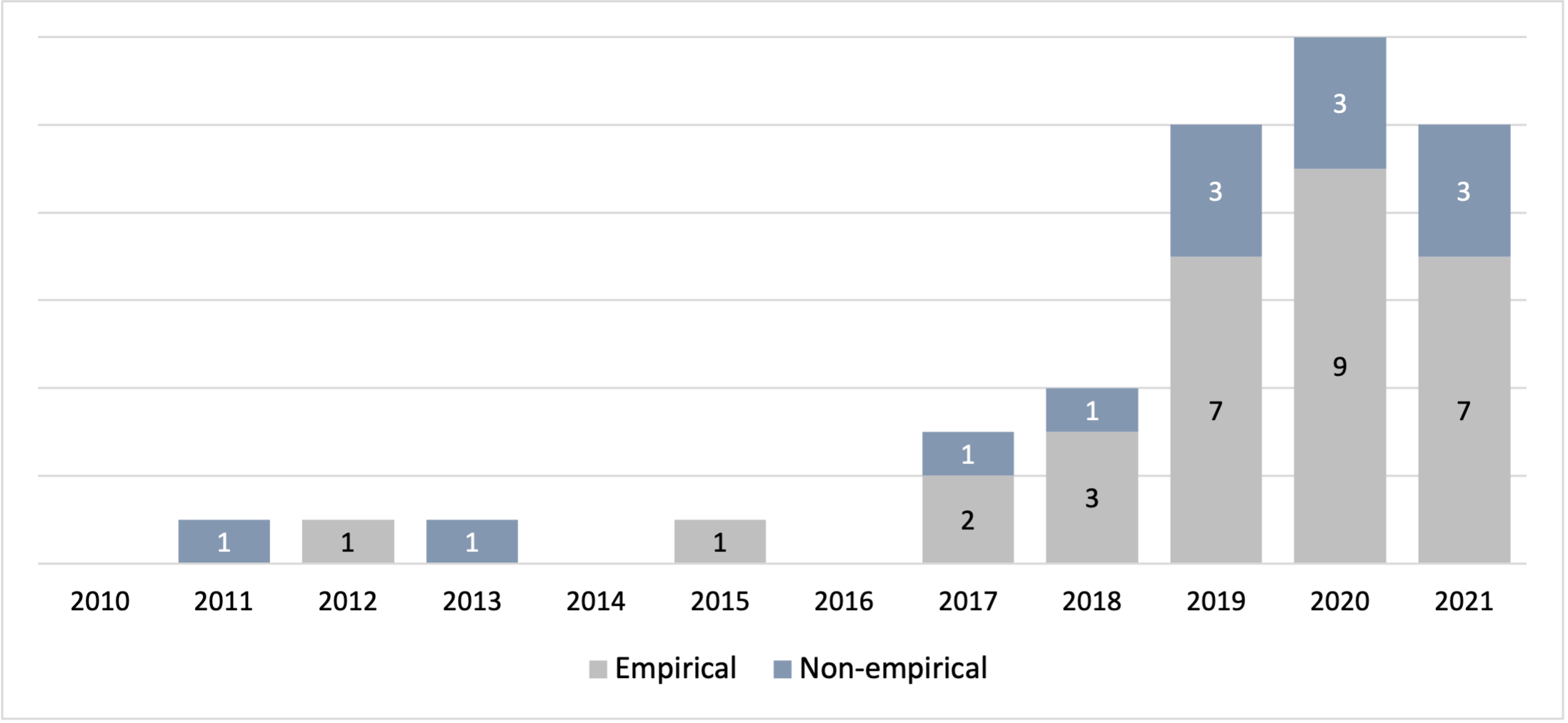}
\caption{Increase shown in the number of publications per year}
\label{fig:PPY}
\end{figure}

\begin{figure}[htbp]
   \centering
\includegraphics[width=0.6\linewidth]{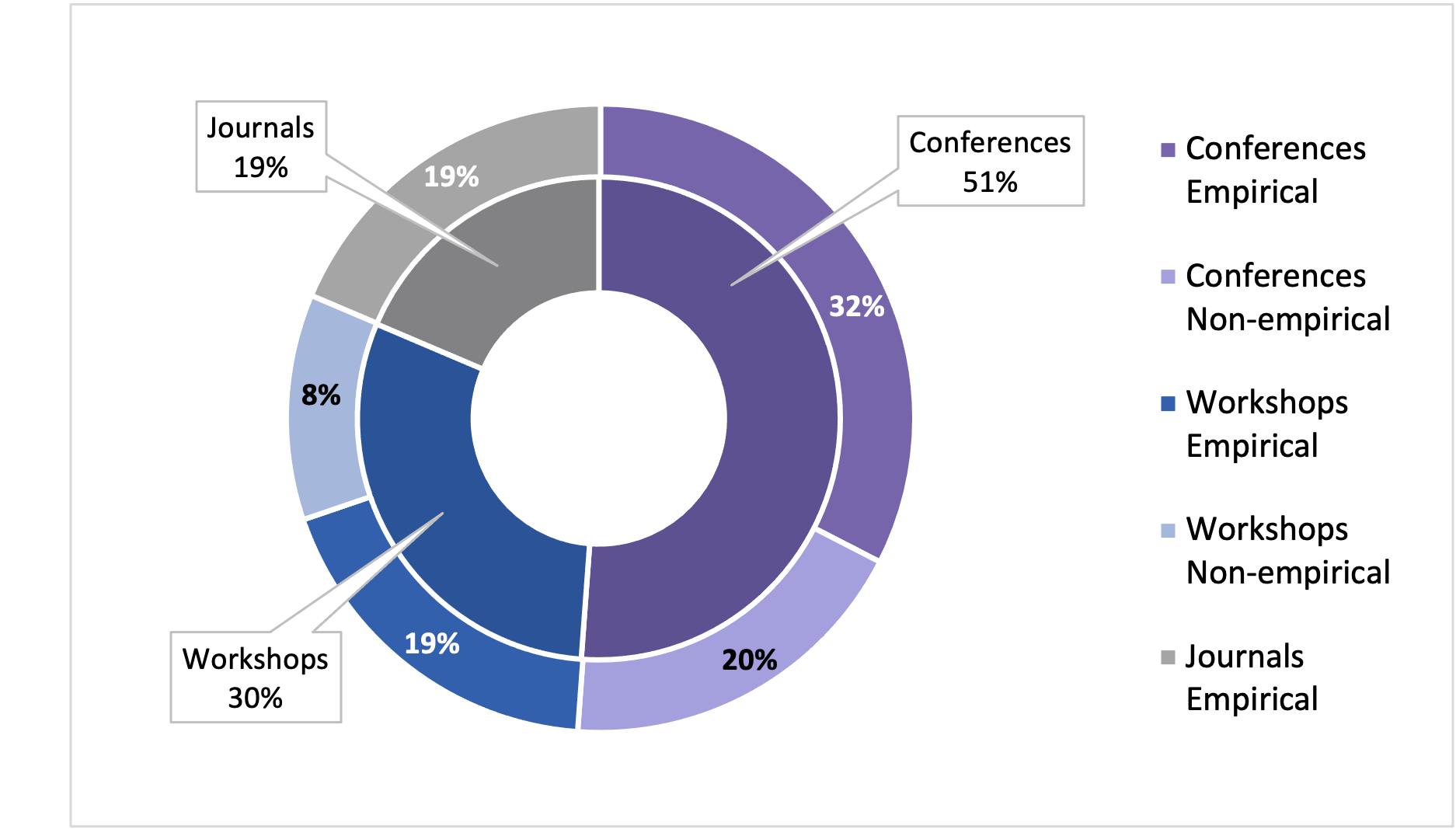}
\caption{Distribution of selected papers at current publication venues}
\label{fig:PublicationTypePercentage}
\end{figure}

Figure.~\ref{fig:PublicationTypePercentage} illustrates the publishing venues for the final papers. Conferences were the most popular place for publishing, with more than half of the published selected primary studies being from conferences.  Workshops were the next popular venue for publications.  Journals were the least popular platform for publications on RE4AI, possibly because the work is relatively new and needs further development to improve its maturity.  Although only eight of the primary papers were published in journals, they were all empirical studies.  Conferences and Workshops had a higher number of non-empirical studies, and more ``idea only" papers (with no empirical evaluation) were presented at such venues.

\subsection{RQ1 Results}

In this section, we provide results from our study for RQ1: \textit{Which Requirements Engineering Frameworks, Notations, Modeling Languages, and Tools have been Proposed to Build AI Systems?}.

\subsubsection{Requirements Engineering Frameworks}

For this research question, we wanted to identify currently used or proposed frameworks to manage and combine RE methods for AI systems.  We identified studies with frameworks if they provided more than one level or layer that build up in providing a holistic solution to manage requirements for AI systems. We only selected frameworks that focused entirely on RE.  For example, in DoReMi \cite{schoonderwoerd2021human}, only one layer of the framework focuses on RE, therefore it was not included as an RE4AI framework. Three different frameworks were identified to include: ``Holistic DevOps" \cite{bosch2018takes} that combined multiple approaches to building AI software. ``Ethics-Aware SE" \cite{aydemir2018roadmap} for analyzing ethical requirements.  And finally, RE4ML \cite{nalchigar2021modeling} that aided in requirements elicitation for ML systems.

\textbf{Holistic DevOps Framework:}  The authors in \cite{bosch2018takes} first identified three different approaches to building software, and a ``Holistic DevOps Framework'' was created to combine all three practices.  The framework sets rules as to when each approach should be used.  The authors mention that some companies currently use this framework in the industry.  However, it is not addressed in research.  These practices included:
\begin{itemize}
    \item \emph{Requirements driven development} can be applied when features are clearly understood, well documented, and used as the basis for building a software system to deliver stakeholders needs.  This approach is suitable for projects that do not require frequent changes to the system~\cite{bosch2018takes}.  Tuncali et al. \cite{tuncali2019requirements} utilized this approach to test requirements in a virtual environment for a self-driving car.  They found that this method reduced the need for resources to design and test the system.
    \item \emph{Data/outcome driven development} experiments with different methods and solutions to achieve the desired outcome.  This approach works with projects that require frequent updates and new features.  The method monitors large amounts of data to find specific patterns and is commonly used in online businesses~\cite{bosch2018takes}.  Design decisions and systems characteristics of data-driven approaches are usually defied by the analysis of recorded data~\cite{bach2017data}.
    \item \emph{AI-driven development} is used when large amounts of data exist but limited resources to manage the data.  Automotive companies with image recognition or user interfaces that use speech recognition tend to use this method.  It is also ideal for predicting future activities from existing patterns found in data \cite{bosch2018takes}.
\end{itemize}

Bosch et al.~\cite{bosch2018takes} explained that businesses are moving towards data-driven approaches, as decisions are becoming more dependent on data to determine the system's functionalities.  This change resulted in the demand to modify current RE practices to become more adaptive to data-driven approaches.  Also, data-driven RE is changing the way requirements are elicited and obtained as the traditional techniques of interviews and questionnaires are changing to include information obtained from social media and online forms~\cite{kostova2020interplay}.



\textbf{Ethics-Aware SE:} Aydemir and Dalpiaz \cite{aydemir2018roadmap} proposed a method that would allow requirements engineers and stakeholders in analyzing ethical requirements.  Ethics-aware SE consisted of five phases to include:

\begin{itemize}
    \item Articulation:  This phase involved eliciting and modeling ethical requirements from stakeholder while ensuring that both parties were inline with the ethical values for the proposed system. Ethical values could include diversity, ensuring user privacy, transparency, work ethics, etc.
    \item Specification: Involves matching the ethical requirements identified in the first step to the systems functional and quality requirements.
    \item Implementation:  Applying the ethical requirement and building the software product.
    \item Verification: Continuously checking if the product is inline with the ethical requirements.
    \item Validation: Testing if final software product is aligned with the ethical requirements proposed by the stakeholders in the first phase.
\end{itemize}

Each phase involved a method to extract, manage or evaluate ethical requirements.  The authors argued that ethical requirements need to adapt to the changes in today’s software systems.  So providing a platform that allows the stakeholder, developer, and requirements engineer to collaborate during the different RE phases could help mitigate ethical issues. 

\textbf{GR4ML Framework:} Nalchigar et al. \cite{nalchigar2021modeling} presented a conceptual framework that offered three modeling views.  The three views include:

\begin{itemize}
\item Business View:  The first view facilitated addressing and setting requirements related to the business domain and user needs.  In this layer the business related goals are elicited and modeled.  The goals include `decisions' to be made,  `questions' to help make decisions, and `insights' to provide knowledge to the question goals.
\item Analytics Design View:  The second view involved the selection of technical features such as machine learning algorithms and what qualities and trade-offs should be considered when choosing an algorithm.  Analytic had three main goals to include: `prediction' goal that intends to predict a value based on existing attributes, `description' goals for describing the selected dataset, and `prescription' goal to find the best alternative among given options.  The analytic view also included softgoals to capture qualities and an indicator to measures the performance of the softgoal. 
\item Data Preparation View:  The third view assisted in the selection and understanding of available data sets.  This view helped in preparing data to use and share among the team \cite{nalchigar2018business}.
\end{itemize}

All three views were combined and linked together to provide a holistic view of the entire system and its connections.

\subsubsection{Existing Modeling Languages, Requirements Notations found in RE4AI}

In total, 21 studies used modeling languages or requirements notations to present requirements.  The most popular modeling notations and languages among the studies were UML, Domain Specific Modeling (DSM), followed by GORE, as shown in Figure.~\ref{fig:ModelingLanguages}.  The study in~\cite{bonfe2012towards} combined two modeling techniques to produce their model, while ~\cite{gruber2017integrated} created an extension to an existing one.  We found that with the new search (mid-2020 to mid-2021), there was an increase in studies that used DSM.  Five of the added 16 new studies used DSM to display requirements for specific application domains.  Three studies focused on using UML, and no additional studies were found to use GORE.  This shift shows that more studies are growing towards using DSM to present requirements for AI-software applications.

\begin{table}[h!]
\renewcommand{\arraystretch}{1.2}
\caption{Modelling languages and requirements notations used in selected studies} 
\label{table:ModelingLanguages}
\centering
\footnotesize
\begin{tabular}{p{4.5cm} p{4.5cm}}
\hline
\textbf{Modelling language} & \textbf{Study}      \\ \hline
GORE (Goal-Oriented~RE) & \cite{bonfe2012towards, dimitrakopoulos2019alpha, neace2018goal, lockerbie2020using, ishikawa2020evidence} \\ \hline
UML / SysML / Use Cases     & \cite{bonfe2012towards, gruber2017integrated, jakob2017defining, altarturi2017requirement, amaral2020ontology, ries2021mde, cleland2020requirements, schoonderwoerd2021human} \\ \hline
Signal Temporal Logic (STL) & \cite{tuncali2019requirements} \\ \hline
Trafﬁc Sequence Charts (TSC)     & \cite{becker2020partial}\\ \hline
Casual Diagrams & \cite{schwammberger2021quest}\\
\hline
Domain Specific Models          & 
\cite{weihrauch2018conceptual, olmos2020helping, agrawal2020model, nakamichi2020requirements, islam2021mobile, hall2019systematic}\\ \hline
\end{tabular}
\end{table}

\begin{figure}[h]
   \centering
\includegraphics[width=0.6\linewidth]{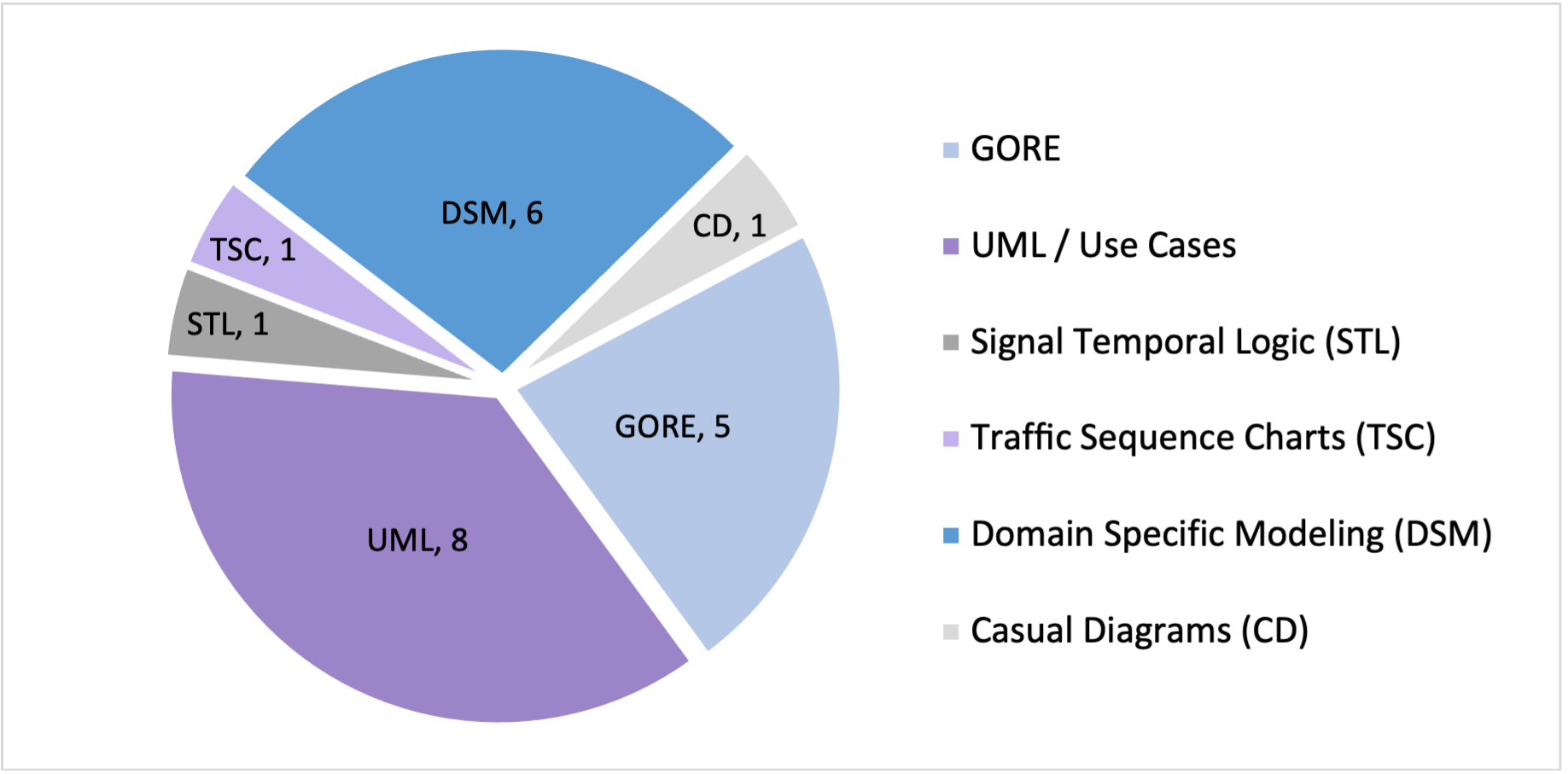}
\caption{The distribution of modeling languages and requirements notations used in selected studies}
\label{fig:ModelingLanguages}
\end{figure}

\textbf{GORE:} Five studies preferred the use of goal-modeling techniques.  The authors in \cite{bonfe2012towards} argued that using goals to present requirements for surgical tasks was more favorable as they seemed to be in line with how surgeons think or perceive tasks from a medical perspective.   Dimitrakopoulos et al.~\cite{dimitrakopoulos2019alpha} stated that goal-oriented methods were more suited for capturing business requirements.  Neace et al.~\cite{neace2018goal}  chose to model with GRL because it provided better support for modeling non-functional and quality requirements.  They also stated that GORE has become more popular in modeling requirements for autonomous systems.  Ishikawa et al.~\cite{ishikawa2020evidence} proposed to use GORE-MLOps, a methodology that adapts requirements analysis from GORE methods to ML systems.  Finally,~\cite{lockerbie2020using} favored the use of soft goals in i* models as it provided a better description of a person's objective, such as quality of life.   Table~\ref{table:GORE} shows the different studies that used GORE as a modeling language.

\begin{table}[htbp]
\renewcommand{\arraystretch}{1.2}
\centering
\footnotesize
\caption{The different GORE methods presented in selected studies} 
\label{table:GORE}

\begin{tabular}{p{7cm} p{7cm} p{1cm}}
\hline
\textbf{Gore type} & \textbf{Description}  & \textbf{Study}   \\ \hline
FLAGS (Fuzzy Live Adaptive Goals for Self-adaptive systems)~\cite{baresi2010fuzzy}   &  Used to presents requirements for tasks performed by a smart surgical robot & \cite{bonfe2012towards}\\ \hline
CORE (Capability Oriented Requirements Engineering) & Uses goals to capture the systems current and desired capabilities & \cite{dimitrakopoulos2019alpha}  \\ \hline
GRL (Goal-oriented Requirements Language) & Modeled requirements for an autonomous aircraft system to detect radiation levels in disasters & \cite{neace2018goal}  \\ \hline
i* & Created a model for people living with dementia using i* soft goals &   \cite{lockerbie2020using} \\ \hline
GORE-MLOps &  Proposes a methodology to models uncertainty in requirements for AI systems   & \cite{ishikawa2020evidence}\\ \hline
\end{tabular}
\end{table}

\textbf{UML:} Eight papers used UML to model requirements, as shown in Table \ref{table:UML}.  Three of these studies used Use Cases \cite{jakob2017defining, altarturi2017requirement, schoonderwoerd2021human}.  In \cite{gruber2017integrated} SysML is used to model functional and non-functional requirements.  SysML allowed them to graphically present requirements and their relationships, therefore providing a visual view of the behaviors between each requirement.  However, the drawback to using SysML was that it did not provide enough aid to model non-functional requirements.  As a result, the study proposed and tested an extension to SysML to support non-functional requirements.  Amaral et al. \cite{amaral2020ontology} proposed an ontology of trust to help define requirements for trustworthy AI.  The study implements these trustworthiness requirements in an OntoUML model.  The model also assisted in displaying any risks related issues when it came to trust.  In \cite{cleland2020requirements} Activity Diagrams were used to model requirements for a configuration system. And finally, \cite{ries2021mde} used a semi-formal UML model to generate their dataset requirements concept model (DRCM).  As reported by the authors, the rationale behind using UML was the ease of using UML for non-software engineers. 
 
\begin{table}[htbp]
\renewcommand{\arraystretch}{1.2}GORE-MLOps 
\centering
\caption{The different UML methods presented in selected studies} 
\label{table:UML}
\footnotesize
\begin{tabular}{ p{3cm} p{11cm} p{1cm}}
\hline
\textbf{UML type}       & \textbf{Description}  & \textbf{Study}   \\ \hline
Statechart and sequence diagrams  & Uses statecharts and sequence diagrams to model the medical robots procedure, and the interaction between the system and the user  & \cite{bonfe2012towards}\\ \hline
SysML  &   An extension of SysML is used to model functional and non-functional requirements for automotive car systems & \cite{gruber2017integrated}\\ \hline
Use Cases  & First study created use cases for traffic scenarios and proposed navigation solutions for vision impaired people. Second study generated a use case for a young girl suffering from Attention Deficit/- Hyperactivity Disorder (ADHD)  & \cite{jakob2017defining, schoonderwoerd2021human} \\ \hline
Actionable Use Case  &  Proposes an actionable use case diagram as a means of collaboration between the data scientist and the software engineer  & \cite{altarturi2017requirement}\\ \hline
OntoUML &  Proposes an ontology using UML to model trustworthy requirement & \cite{amaral2020ontology} \\ \hline 
Semi-formal UML & Proposed a model called "Dataset Requirements Concept Model" (DRCM) to display requirements needed for the structure of datasets & \cite{ries2021mde} \\ \hline
Activity Diagram & The study gathers requirements and models them using use cases and activity diagrams to model a configuration system for an Unmanned Aerial Vehicles & \cite{cleland2020requirements} \\ \hline
\end{tabular}
\end{table}

\textbf{Signal Temporal Logic:}  Signal Temporal Logic (STL) is a specification language that enables real-time reasoning of properties by providing past, and future variables \cite{bartocci2018STL}.  The study in \cite{tuncali2019requirements} used STL to specify requirements for a perception system in an autonomous vehicle.   It provided features such as reachability, safety, and reactive requirements to include in the specifications.   They then mapped these requirements into three testing scenarios using the Sim-ATAV framework and a virtual environment to generate test cases for autonomous vehicles.

\textbf{Trafﬁc Sequence Charts:}  Trafﬁc Sequence Charts (TSC) is a graphical specification language used for traffic scenarios.  TSC is based on snapshots, and each snapshot represents a traffic situation.  When assembled, snapshot charts consist of history, future, and consequence.  Snapshots are also linked or combined with operations such as sequences and choice.  The work in \cite{becker2020partial} used TSC to display requirements for an autonomous vehicle.  The main objective of the study was to find any inconsistencies in TSC requirements specifications.

\textbf{Causal Diagrams:} Causal diagrams are mathematical graphical notations that represent statistical relationships between objects \cite{glymour2008causal} and are used to identify specific variables or measures. They are often used in epidemiology to minimize biases \cite{pearl1995causal, greenland1999causal}.  In \cite{schwammberger2021quest} the author proposes to use casual diagrams to generate different explanation paths in order to provide a user-understandable explanation for self-explainable systems.  The study uses the framework proposed in \cite{blumreiter2019towards} ``Monitor, Analyse, Build, Explain" (MAB-EX) to provide explanations for the systems crossing controller that monitors traffic intersections.

\textbf{Domain Specific Modeling:} Domain Specific Modeling (DSM) is defined by Fowler as a ``computer programming language of limited expressiveness focused on a particular domain” \cite{fowler2010domain}. Six studies used DSM to model requirements \cite{weihrauch2018conceptual, olmos2020helping, agrawal2020model, nakamichi2020requirements, islam2021mobile, hall2019systematic}.  The study in \cite{weihrauch2018conceptual} builds a model for a Smart Process Control System based on requirements gathered from the industry and survey results.  In \cite{olmos2020helping} the authors used the Knowledge Management on a Systematic process for Requirements Engineering (KMoS-REload) to elicit and model external and internal knowledge using the KMoS-REload process.  In \cite{agrawal2020model} requirements were modeled to show the interaction between humans and a drone to detect hot spots and victims through a fire in a building.  In \cite{islam2021mobile}, a model was built to present the framework for a prototype application that was based on requirements elicited from interviews for a mobile health application.  And finally, Hall et al. \cite{hall2019systematic} built a model for explainable AI that was then applied to an industrial case study.

\subsubsection{Modeling Tools}

We found the use of two different tools.  The {jUCMNav} tool was used in two studies and is a free graphical editor that supports modeling requirements in Goal-oriented Requirement Language (GRL) and Use Case Maps (UCM) \cite{amyot2011grl}.  Neace et al.
\cite{neace2018goal} implemented the requirements proposed in \cite{vassev2014autonomy} using the goal-oriented requirements language GRL. These requirements were then modeled using the jUCMNav tool to present them graphically.  In \cite{ishikawa2020evidence} the study used jUCMNav to present their goal model that represented uncertainty in AI systems.  The second tool was created by Ries et al.~\cite{ries2021mde} who developed a toolset based on the {Sirius framework} to present their modeling language visually.  The Sirius framework is an open-source graphical editor that allows for domain-specific models to be presented visually~\cite{viyovic2014sirius}.

\subsection{RQ2 Results}
In this section, we discuss results pertaining RQ2: \textit{Which evaluation methods used to assess empirical studies in RE4AI?}
We identified three types of empirical investigations used in evaluating existing RE4AI techniques, models, and frameworks. These methods included case studies, experiments, and surveys.  Figure.~\ref{fig:MethodDistribution} shows that the most common type of methods used were case studies followed by experiments.  We also obtained the different data collection methods used in each empirical evaluation as shown in Figure.~\ref{fig:datacollection}.  We found interviews to be the most popular form of data collection, followed by using existing or creating new databases.  Other data collection forms included workshops, meetings, focus groups, gathering data from online websites and user feedback, etc.

\begin{figure}[htbp]
   \centering
\includegraphics[width=0.6\linewidth]{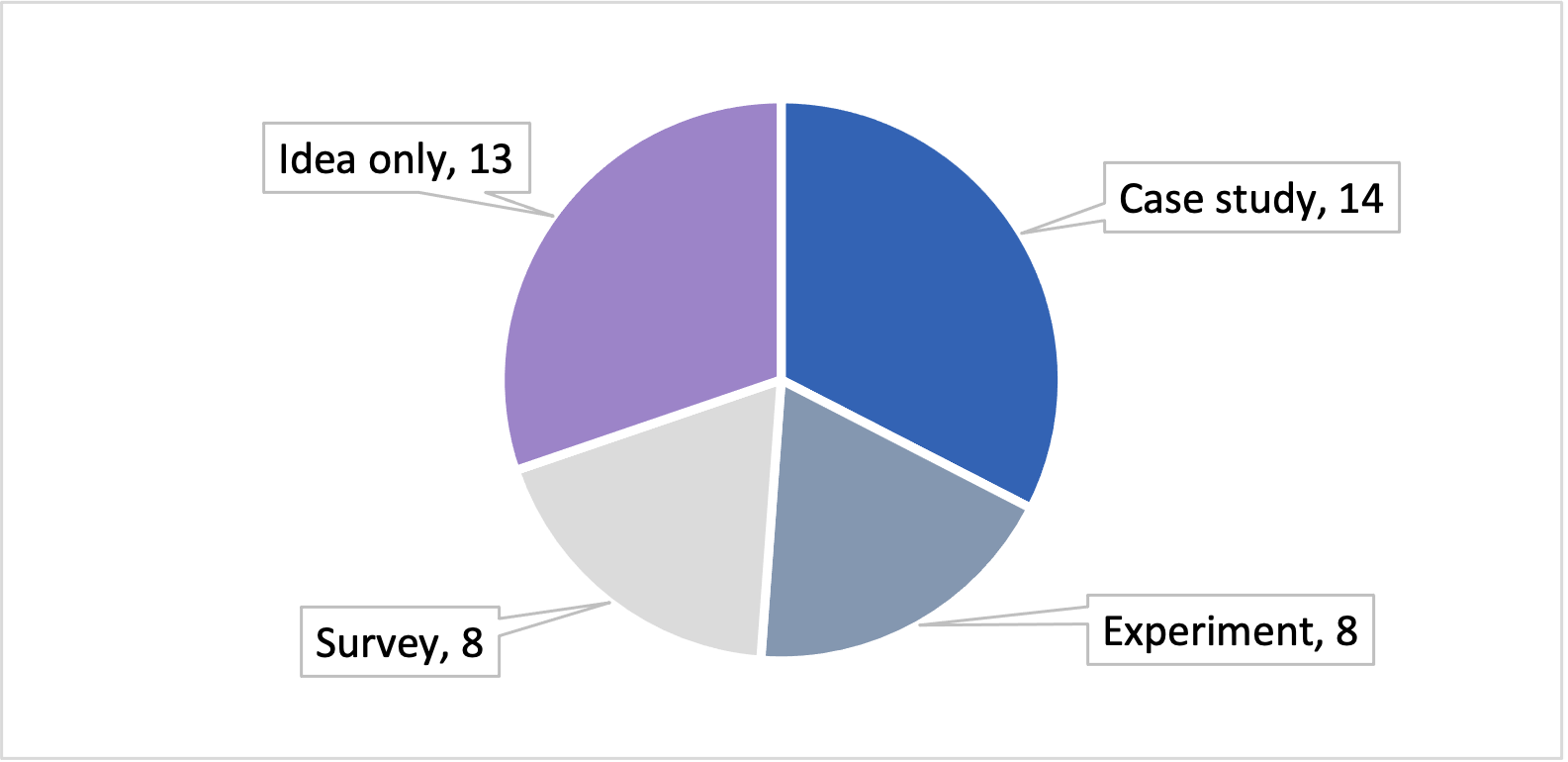}
\caption{The distribution of the methods found in the selected primary studies}
\label{fig:MethodDistribution}
\end{figure}

\begin{figure}[htbp]
   \centering
\includegraphics[width=0.9\linewidth]{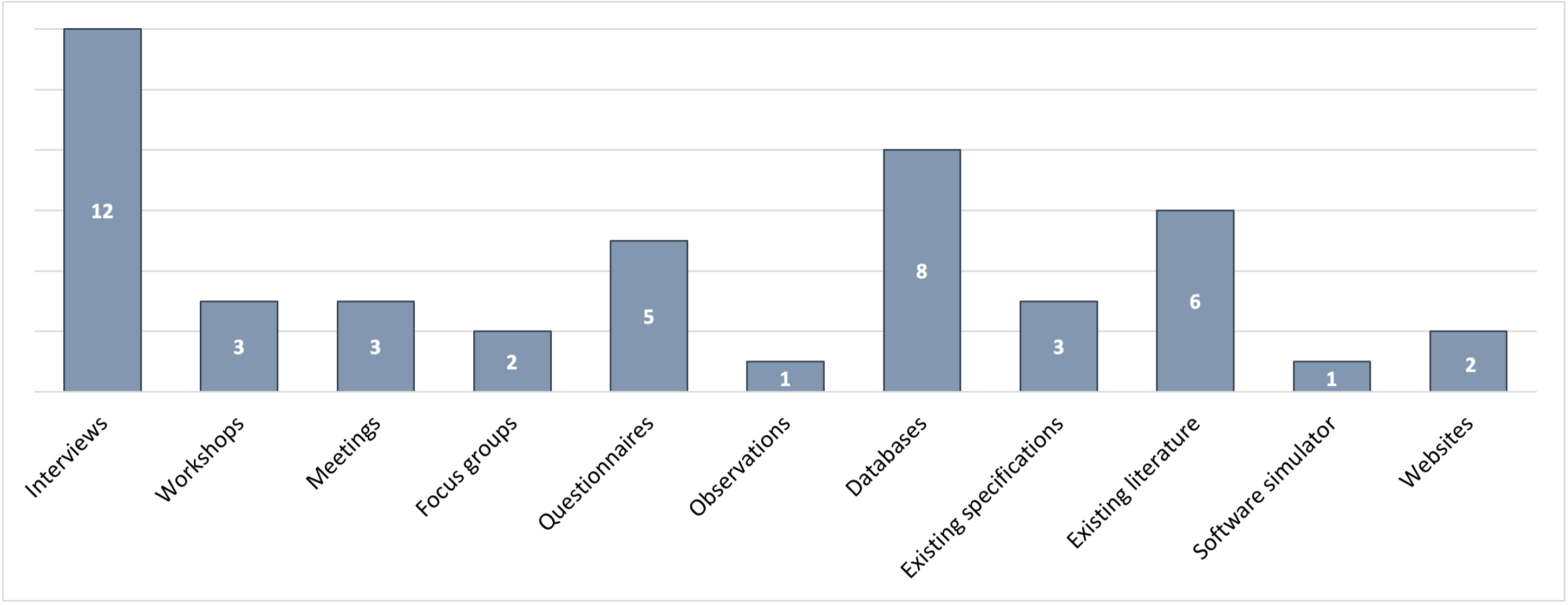}
\caption{The different data collection methods used in selected empirical evaluations and how often they were used}
\label{fig:datacollection}
\end{figure}

\begin{table}[]
\caption{Papers with Case Study Research} 
\label{table:CaseStudies}
\centering
\footnotesize
\begin{tabular}{p{0.4cm} p{5cm} p{5cm} p{3cm}}
\hline
\textbf{Study} & \textbf{Case Specification} & \textbf{Outcomes / Aims} & \textbf{Data Collection} \\ \hline

\cite{gruber2017integrated} & An Active Lane Change Assistant, to support drivers when changing lanes in autonomous cars & Better connectivity and requirements tracing and was easier to track and implement changes & Existing specification sheets \\ \hline
\cite{sandkuhl2019putting} & Feasibility of organizations in using AI solutions and for a banking systems fraud detection software & The case did not have enough training data on fraud cases to use AI as a solution & Interviews \\ \hline
\cite{rahimi2019toward} & Finding requirements for ``pedestrians" for a self-driving & A model that displays all requirements needed to describe a pedestrian &  Web-search engines + ``Caltech" dataset \\ \hline
\cite{bosch2018takes} & Identify different approaches when building AI software &  Proposed a ``Holistic DevOps" Framework & Meetings, interviews, workshops \\ \hline
\cite{fenn2016addressing}& Addresses three NFR's for a robots vision system for a RoboCup competition & Results showed better performance in the new vision system & A dataset with 5 pre-recorded image streams \\ \hline
\cite{neace2018goal} &  Implement requirements for an autonomous aircraft system to detect radiation levels in disasters  & Aimed to optimize high-performance, maximize safe operations and low cost    & Requirements from \cite{vassev2013autonomy} +  Published data   \\ \hline
\cite{nakamichi2020requirements} & Evaluate the effectiveness of the quality characteristics for ML systems & Found that 22 of the issues were present in the requirements definition & Surveys, prior research, collaborations   \\ \hline
\cite{challa2020faulty} & Feeding faulty data into a DL system for a wastewater plant and monitors the behavior &  Identifies characteristics needed to improve the quality of data requirements & Data collected from an overflow sewer site \\ \hline
\cite{ries2021mde} & Identifying images of digits between 0 to 9 using a DL model  & Improved quality of datasets and identify  issues with digits early & Images of five-segments ``characterized digits"\\ \hline
\cite{olmos2020helping} & Update a traffic monitoring and routes management systems for a freight company using AI & Establish functional requirements and improved teams communications & Focus groups \\ \hline
\cite{ntakolia2020user} & Eliciting Requirements for a navigation system for vision-impaired people and the prototype is tested & Collected requirements could have potential benefits to developing user-centric SAS &  Literature, Interviews, questionnaires, focus groups \\ \hline  \cite{nalchigar2021modeling}& Evaluates the “GR4ML” framework on a start-up company in healthcare to elicit requirements & New ML requirements are established, and improved team communication & Interviews, meetings \\ \hline
\cite{schoonderwoerd2021human}& Gathering requirements from pediatricians for a young girl suffering from ADHD & Used to evaluate the method called “DoReMi” for AI-generated explanations & Questionnaire and interviews\\ \hline
\cite{islam2021mobile} & Prototype for a mobile application to identify if the user had any signs of depression and anxiety  & Initial prototype had potential to aid with people’s mental wellbeing during the Covid-19 lockdown  & Interview 37 people and 20 participated in a survey \\ \hline
\end{tabular}
\end{table}

\subsubsection{Case Studies}

We identified 14 studies that performed a case study.  Table~\ref{table:CaseStudies} display's the list of all case studies found and the outcome of each one.  The authors in ~\cite{gruber2017integrated, neace2018goal, olmos2020helping} conducted case studies to evaluate a model that was created based on a set of given requirements.  Nakamichi et al~\cite{nakamichi2020requirements} propose a quality evaluation model and a method to measure functional correctness for ML systems, and three case studies were carried out to evaluate the proposed model.  In~\cite{sandkuhl2019putting} the results justified what information was needed and what was missing to implement AI as a software solution for organizations.  In their case, they found that they did not have enough training data to implement AI as a solution.  The authors in~\cite{bosch2018takes} conducted several case studies with the industry to identify three different approaches to building AI-based software systems and proposed a ``Holistic DevOps Framework''.  

Fenn et al.~\cite{fenn2016addressing} evaluate a new architecture design for a vision system using a case study.  The new architecture focused on three non-functional requirements: portability, extensibility, and modifiability, and the authors found the system performed better with the new design.   In ~\cite{islam2021mobile} requirements are elicited for a mobile health application and the feasibility of these requirements are evaluated.  Rahimi et al.~\cite{rahimi2019toward} worked on finding specifications for machine learning components on safety-critical domains.  The study proposed a solution to extract and visualize domain specifications for requirements.  They focused on extracting requirements for ``pedestrians” and how a self-driving car would recognize~pedestrians.  

Challa et al.~\cite{challa2020faulty} proposed characteristics needed for data requirements to provide quality data for deep learning systems.  In~\cite{ntakolia2020user} the study elicits requirements for a smart assistive system and creates a prototype to validate and test the effectiveness of these requirements.   Ries et al. \cite{ries2021mde} used a case study to improve data requirements for Deep Learning (DL) systems.  The study identified requirements needed for the database’s structure.  And finally, in \cite{schoonderwoerd2021human} a method called “DoReMi” is built for AI-generated explanations.   A case study is conducted on a clinical trial for a young girl suffering from Attention-Deficit/- Hyperactivity Disorder (ADHD) to evaluate explanations generated by the AI system.

\subsubsection{Experiments}
 
Eight studies used experiments to evaluate their proposed solutions as shown in Table~\ref{table:Experiments}.  Three studies generated experiments to evaluate a model or method presented~\cite{becker2020partial, dimitrakopoulos2019alpha, cleland2020requirements}.  Another two studies experimented on finding data requirements.  In~\cite{shin2019data} the study focused on finding data requirements for monitoring energy consumption for houses in Japan, and~\cite{bonfe2012towards} gathered data to use for training a surgical robot.  In the last three experiments,  ~\cite{hu2020towards} evaluates the robustness of requirements for a computer vision system, ~\cite{tuncali2019requirements} tests requirements for a self-driving car, and \cite{samin2021towards} experiments on the threshold needed to find the trade-off on the impact of a NFR on the rest.

\begin{table}[]
\caption{Papers with Experiments Research} 
\label{table:Experiments}
\centering
\footnotesize
\begin{tabular}{p{0.4cm} p{5cm} p{5cm} p{3cm}}
\hline
\textbf{Study} & \textbf{Experiment} & \textbf{Outcomes / Aims} & \textbf{Data Collection} \\ \hline
\cite{bonfe2012towards} &  Requirements for a surgical robot to find the force required when inserting and removing the needle into cancerous cells &  Data collected is to be used in a future case study to train a robot to perform a procedure on removing cancerous cells from a kidney & Interviews + manually collecting force to insert and remove a needle \\ \hline
\cite{hu2020towards} &  Adding noise to a vision detection system (YOLO) to identify images for a self-driving car & Using YOLO is not robust against changes and could have severe consequences in safety-critical systems  & Image from Berkeley DeepDrive \cite{yu2020bdd100k} and Gaussian noise \\ \hline
\cite{tuncali2019requirements} & Requirements are used to build a virtual environment to test functionalities of a self-driving car & Capturing unacceptable behaviors and identify the test cases that violate the requirements & Data gathered from 3 sensors:  CCD camera, lidar, and radar\\ \hline
\cite{becker2020partial} & Prototype tool to maintain requirements consistency for autonomous driving and analyze a traffic situation & They could not identify all conflicts and requires further research to create refined consistency notations & Results taken from previous studies\\ \hline
\cite{shin2019data} & Implementing different algorithms on a dataset to find the effects of sampling rate and house numbers 
& Algorithms such as classification and regression performed badly when the sampling rate was low & Data on energy consumption for 3 appliances and 58 houses \\ \hline
\cite{dimitrakopoulos2019alpha} & Capabilities are presented in a goal model and tested in a traffic simulation to demonstrate its efficiency  & The method could be effective in the assessment of real-time traffic situations  &  Software simulator that generated traffic scenarios \\ \hline
\cite{samin2021towards} & Satisfaction of NFR - the threshold is calculated based on the trade-off on the impact of a NFR on the rest & The outcomes showed improved results when compared to other methods used to prioritize requirements & Existing requirements specifications \\ \hline
\cite{cleland2020requirements} & Requirement driven approach is used to create a configuration system for a product line & The study gathers requirements and models them using use cases and activity diagrams & Literature and Existing requirements \\ \hline
\end{tabular}

\end{table}

\subsubsection{Surveys / Interviews}

Survey methods were used as means to gather data for both case studies and experiments.  As shown in Tables~\ref{table:CaseStudies} and~\ref{table:Experiments}, several studies used surveys to collect data either to build an artefact or to evaluate it.  However, we found seven studies that explicitly used surveys as an evaluation technique.  These techniques included questionnaires, interviews, workshops, and observations.  Table~\ref{table:Interviews} shows the studies that only performed surveys along with the outcomes.  In~\cite{jakob2017defining} the study uses interviews to gather requirements for an orientation and navigation system that is based on computer vision for people who are vision impaired.  Lockerbie et al.~\cite{lockerbie2020using} conducts workshops to identify soft goals for an AI application that would provide better quality of life for people with dementia.  In \cite{rivero2021lessons} the study carried out interviews to identify requirements for a system that flags users with the potential of filing a lawsuit against a given power company.  The system also provided information on these customers to the company’s lawyers and approaches such customers to discuss available solutions through a chatbot function.   Both studies in ~\cite{vogelsang2019requirements, habibullah2021non}  conduct interviews to shed light on the challenges RE faces when building ML systems.  And in \cite{weihrauch2018conceptual} the authors evaluate a model for a Smart Process Control System through several questionnaires and workshops.  Finally, both \cite{hall2019systematic, cirqueira2020scenario} use interviews to find requirements for explainable AI.  

\begin{table}[]
\renewcommand{\arraystretch}{1.2}
\caption{Studies using interviews / surveys} 
\label{table:Interviews}
\centering
\footnotesize
\begin{tabular}{p{0.4cm} p{5cm} p{5cm} p{3cm}}
\hline
\textbf{Study} & \textbf{Survey Specification} & \textbf{Outcomes / Aims}& \textbf{Method}  \\ \hline
\cite{jakob2017defining} & Extract requirements, and use cases for six traffic scenarios for a computer vision system & Proposes the use of ADAS algorithms and provide solutions to how they can detect objects & Interviews \\ \hline
\cite{vogelsang2019requirements} & Identify data scientists role in writing, eliciting, documenting and analyze requirements when building ML systems &  The author addresses these challenges by outlining a process for RE4AI systems based on the traditional RE activities & Interviews with 4 data scientists   \\ \hline
\cite{weihrauch2018conceptual} & Identifies functional requirements and performs surveys to evaluate the implementation of these requirements & Identifies an improved set of requirements and a set of four features are specified to build a conceptual model  & Questionnaires, workshops and observations  \\ \hline
\cite{lockerbie2020using} &  An initial model for quality of life is evaluated. Goals were merged and new ones identified & Proposal for an explainable AI app to provide better quality of life for people with dementia  & Workshops with 12 experienced care workers \\ \hline
\cite{cirqueira2020scenario} &  “Scenario-based Elicitation” method is used to identify requirements for explainable AI in fraud detection  & Two scenarios of fraud detection are extracted to be used in future experimental prototypes & Interviews 3 fraud experts working at a bank \\ \hline
\cite{hall2019systematic} & Interviews with 12 experienced aerospace engineers & Identified explainability characteristics to help map requirements to explainable techniques & Formal interviews and informal discussions \\ \hline
\cite{habibullah2021non} & Interviews people from industry to find challenges related to NFR for ML & Identifies the important and less important NFR in ML systems & Interviews ten ML specialists \\ \hline
\cite{rivero2021lessons} & Identify requirements for building a system that identifies power users that have potential to file a lawsuit & Provide information to the company’s lawyers and use a chatbot to approach these customers and discuss available solutions  & Interviews, complaint websites, Power companies laws and rules \\ \hline
\end{tabular}
\end{table}

\subsection{RQ3 Results}

In this section, we elaborate on results for RQ3: \textit{Which target application domains and AI areas of focus considered in the existing approaches?}. The application domain varied between the studies, as shown in Figure.~\ref{fig:ApplicationDomain}. 15 papers listed under general did not specify a domain when applying their concepts. For example,~\cite{horkoff2019nonFunctional, habibullah2021non}, focused on issues and challenges when applying non-functional requirements to AI and ML. Nakamichi et al.~\cite{nakamichi2020requirements} tried to find ways to improve RE techniques for AI systems. They conducted a study to evaluate quality requirements and deliver customer's needs. Bosch et al.~\cite{bosch2018takes} identified the different approaches to building AI systems in general and proposed a framework to combine them. And Ishikawa et al.~\cite{ishikawa2020evidence} presented a model to show uncertainty in AI systems and its impacts on current RE techniques.   Our results showed that the domain with the highest interest in RE4AI were autonomous systems, and most studies in the field of autonomous systems were based on empirical investigations. The following most popular application domains in RE4AI research were medical and assistive technology, and similar to autonomous systems, most of these studies were empirical investigation. 

\begin{figure}[h]
  \centering
\includegraphics[width=0.9\linewidth]{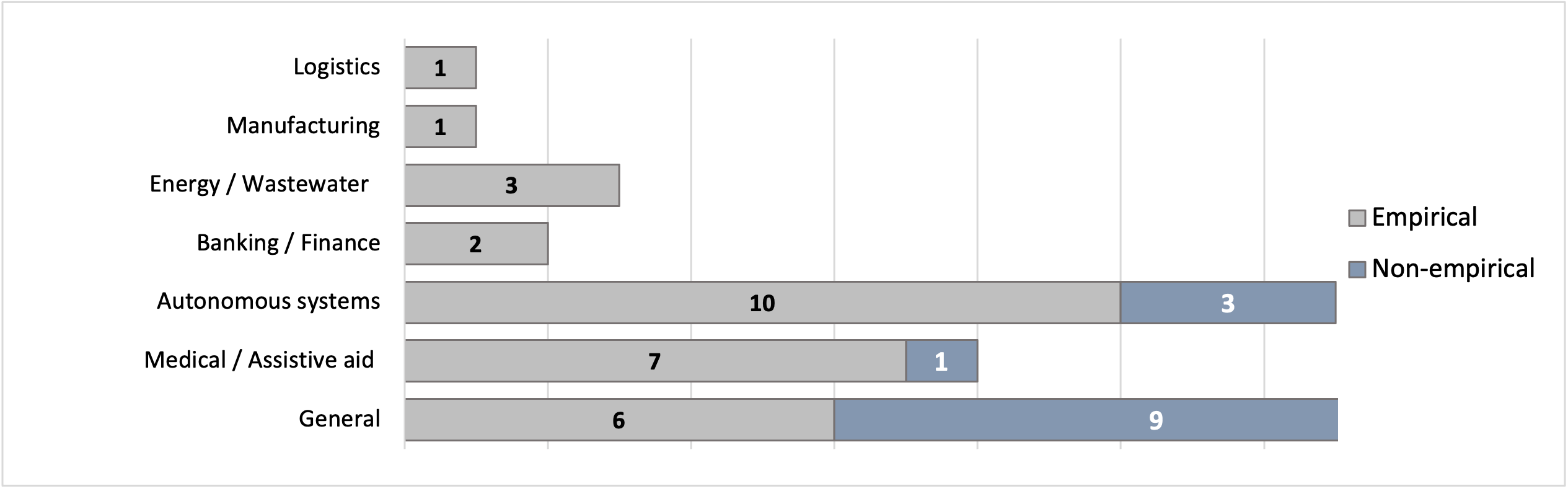}
\caption {Number of studies found in each application domain in RE for AI systems}
\label{fig:ApplicationDomain}
\end{figure}

Figure.~\ref{fig:RE_Domain} shows the different requirements focus domains. Investigating data requirements seemed to gain the most attention.   We found that all empirical studies that involved investigating data requirements were conducted during the past three years (2019, 2020, and 2021), showing that this is an emerging topic and needs further exploration. Several studies have also emphasized the importance of managing data requirements in building AI systems as ``data replaces code"~\cite{vogelsang2019requirements}.   The next most popular research areas were explainability and Non-Functional Requirement (NFR). Some studies reported explainability as a NFR. However, in our classification, we noted studies that only focused on explainability as a separate entity. Studies on explainability were only found after the year 2019. Also, domains that did not have an empirical investigations (non-empirical) were ethics and trustworthy AI as they appeared to be theoretical papers and proposed methodologies that were not yet evaluated.

\begin{figure}[h]
  \centering
\includegraphics[width=0.9\linewidth]{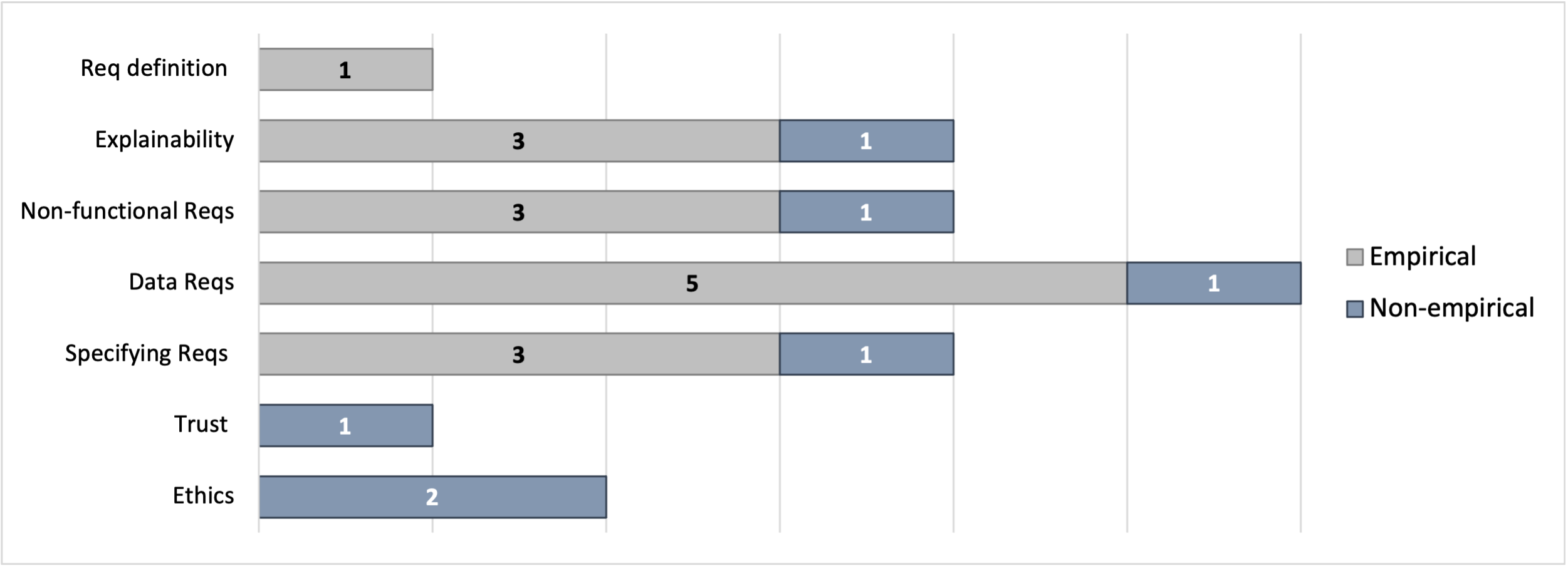}
\caption {The different requirements focus domains addressed in the selected studies}
\label{fig:RE_Domain}
\end{figure}

When looking at the different AI technique used in RE4AI research, we found that most of the studies worked with ML and computer vision as shown in Figure.~\ref{fig:AI_Domain}. Only one study proposes to use Natural Language Processing (NLP)  and presents requirements for a chatbot system \cite{khatamino2021nlp}. Moreover, we did identify a high number of studies that used NLP from our initial search results. However, these studies investigated the use of AI to manage~RE.

\begin{figure}[h]
  \centering
\includegraphics[width=0.9\linewidth]{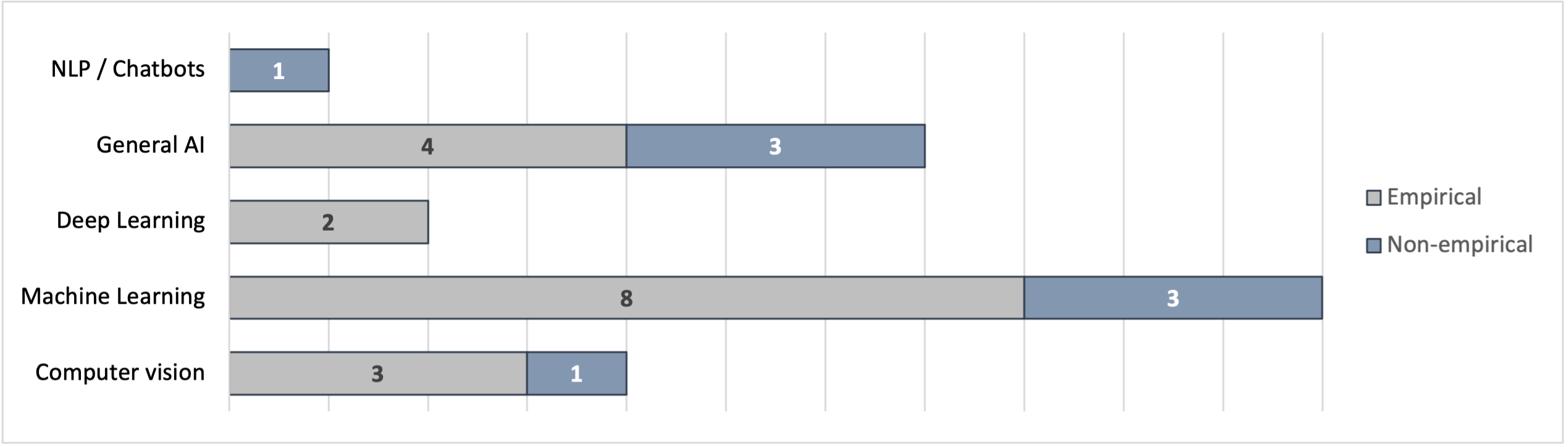}
\caption {The AI topics used in the selected studies}
\label{fig:AI_Domain}
\end{figure}

In Figure.~\ref{fig:DomainVsMethod} we mapped all the application domains against the methodologies used to evaluate empirical work. We found that most of the non-empirical studies were not linked to a specific application domain, thus listed under the general domain.   Some of the studies that focused on autonomous systems used case studies to evaluate their methodologies. However, a more significant number performed experiments. In contrast, medical and assistive technology studies preferred case studies.

\begin{figure}[htbp]
  \centering
\includegraphics[width=\linewidth]{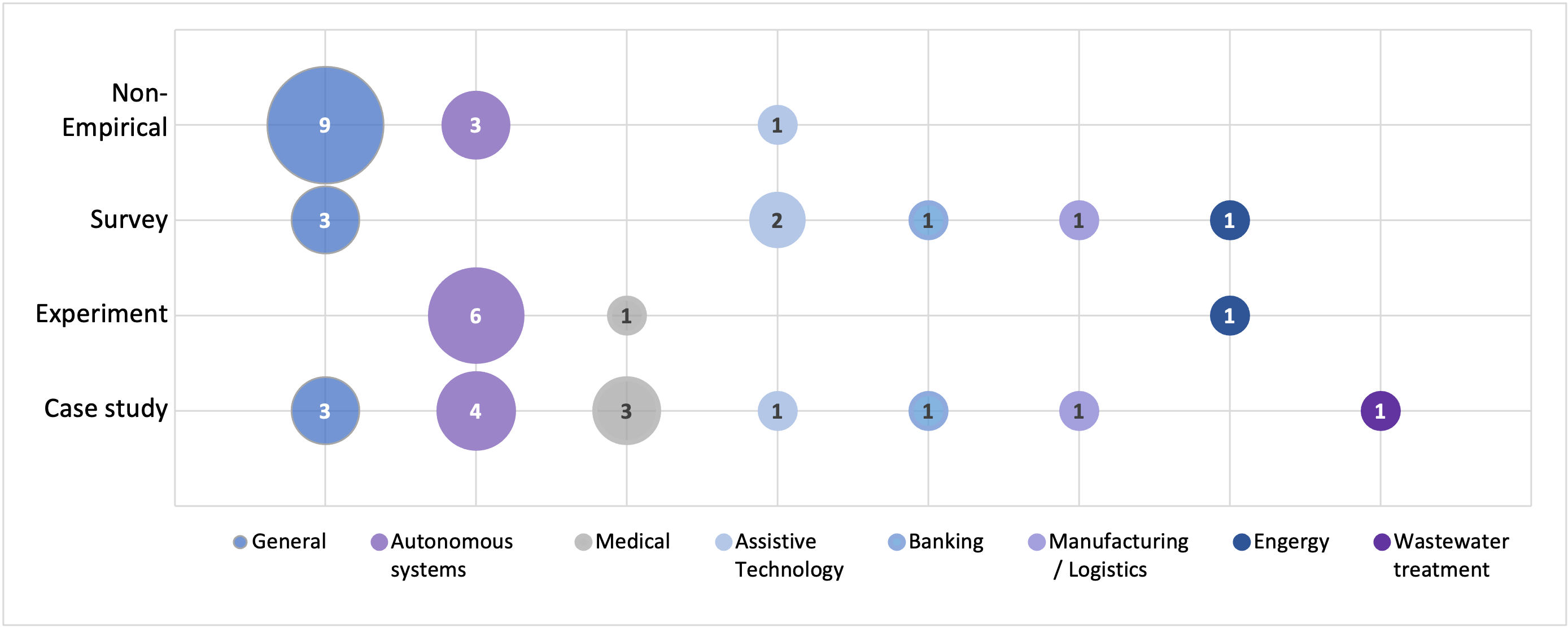}
\caption{Evaluation methodology distribution for each application domain}
\label{fig:DomainVsMethod}
\end{figure}

\subsection{RQ4 Results}
This section discussed our study's results for RQ4: \textit{What are the limitations and challenges of existing requirements engineering techniques when applied to AI systems?}.
We identified challenges and issues presented in the selected primary studies that have emerged due to the shift in RE4AI.  Figure.~\ref{fig:Issues} displays the recurrences of each issue in the literature.  Issues that appeared more often were linked to data requirements and deciding on the trade-off, followed by the emergence of new requirements.  We found that this research question did not change significantly with the additional results obtained from the second search (between mid-2020 and mid-2021).  The observation was that most added studies tended to focus on evaluating proposed solutions rather than presenting issues and challenges. 

\begin{figure}[h!]
   \centering
\includegraphics[width=1\linewidth]{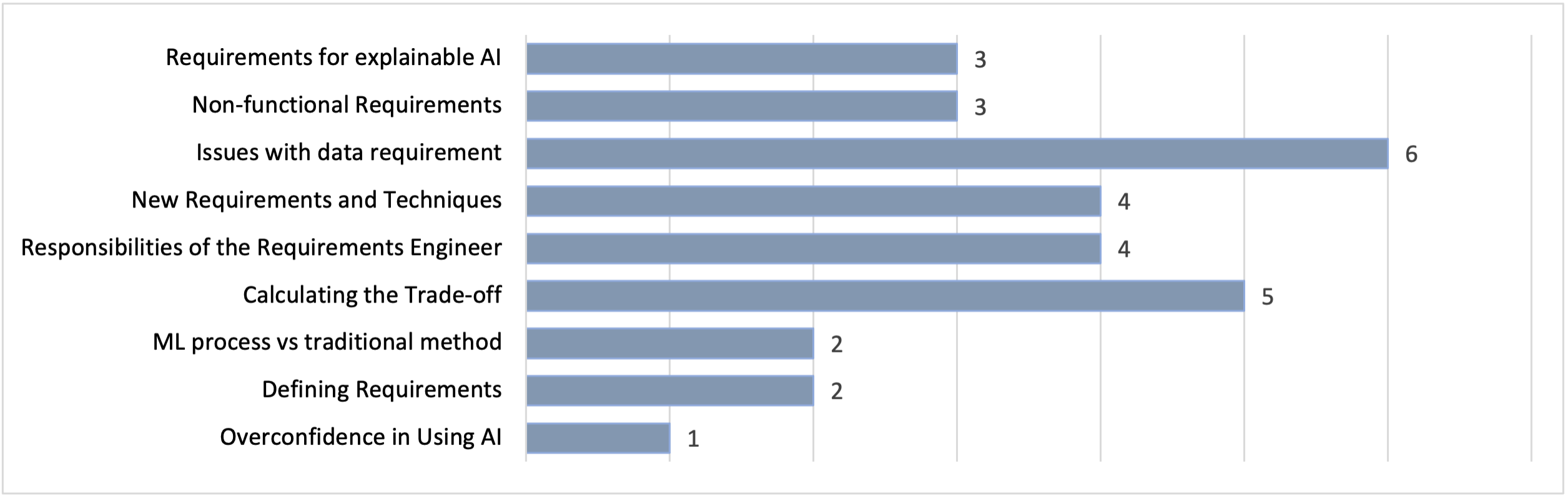}
\caption{Number of RE4AI issues appearing in selected primary studies}
\label{fig:Issues}
\end{figure}

\textbf{Overconfidence in Using AI:}

There is sometimes a  misconception that ``AI will solve everything" for organizations \cite{dimatteo2020requirements}.  Sandkuhl~\cite{sandkuhl2019putting} explained that the general public usually overestimates the capabilities of AI solutions.  During a series of workshops and meetings with several industrial partners, they found that many organizations would choose to use AI without having the experience and expertise in building systems with AI components.  Sandkuhl emphasized that capturing requirements early on in the project is essential so stakeholders clearly understand the capabilities and limitations of AI.  And while AI is not always a feasible solution, companies should establish the need to use AI before proceeding further in the projects. 

\textbf{Defining Requirements:}
Defining requirements for AI systems can be somewhat challenging.  Some requirements might be vague or hard to define.  For example, how do we define a ``Pedestrian" to a self-driving car?  Every person might have a different definition for a pedestrian.  Rahimi et al.~\cite{rahimi2019toward} focused on finding the requirements for ``pedestrians" and how a self-driving vehicle would recognize pedestrians?  The process involved searching for any feature that represented a pedestrian.  Horkoff~\cite{horkoff2019nonFunctional} explained that our understanding of non-functional requirements is not complete, and we need to set out standards to how we define them; for example, how do we define fairness? 

\textbf{Nature of Machine Learning Systems and the Traditional Approach of RE:}  

Traditional non-AI systems are usually deterministic in nature and have a process for RE techniques that are well researched.  However, this is not the case in AI-based software, as outcomes can be unexpected.  In \cite{nakamichi2020requirements} the study emphasized the need to implement new quality techniques for ML systems.  Vogelsang and Borg~\cite{vogelsang2019requirements} explained that the existing methods used in RE need to change to accommodate the different activities currently used for AI systems.  The author defined new types of methods for RE to be included when creating requirements for ML systems.  For example, the elicitation phase should identify requirements, such as data and explainability.  Hence, the need to develop new tools or re-evaluate existing ones to support~RE4AI. 

\textbf{Calculating the Trade-off}

The issue with deciding how to calculate the trade-off came up in 5 different studies.  Horkoff~\cite{horkoff2019nonFunctional} explained that one of the challenges with ML systems' non-functional requirements was to calculate trade-offs when choosing an ML algorithm.  For instance, do we trade privacy for transparency or fairness for accuracy?  And how would we specify or express these choices?  How do we decide on what requirements could be traded and at what cost?  Calculating the trade-off between the impact an NFR will have on the rest of the requirements is also imperative, whether positive or negative.  In \cite{samin2021towards}  a satisfaction threshold is established for the trade-off of NFRs.

The authors in \cite{schoonderwoerd2021human} highlighted the importance of calculating the trade-off between sensitivity and specificity in clinical studies, as clinicians always stress the importance of reducing the number of false negatives when diagnosing patients.  Thus, missing a diagnosis can lead to negative consequences.  In \cite{fenn2016addressing}, the authors traded a slight cutback to efficiency for a substantial increase in modifiability to the system's design.  This cutback provided a more straightforward method to maintain data validity.  Although there was a slight reduction in efficiency, it did not affect the system performance and was still within the processing power range.  Therefore accuracy and reliability remained~unchanged. 

Shin et al.~\cite{shin2019data} worked on finding data requirements for monitoring energy consumption in houses in Japan.  They focused on 2 data requirements: sampling rate and number of samples used.  The study involved collecting data on energy consumption for three different appliances from 58 houses.  They found that it was essential to measure the trade-off between performance/cost when it came to the data's sampling rate.  Algorithms such as classification and regression performed poorly when the sampling rate was low.  So the higher the sampling rate, the better the quality of data was.  The number of houses used was also important when it came to better performance, and including more samples in the dataset provided a more comprehensive range of diversity.  However, increasing the samples could be costly.  Therefore, to what extent can we invest in cost?  There should be a limit to how far we can choose between performance vs. cost.  

\textbf{Responsibilities of the Requirements Engineer:}  Vogelsang and Borg~\cite{vogelsang2019requirements} stated that data scientists are responsible for writing requirements in current ML systems.  As AI and ML are integrating into software systems,  a new role for data scientists is emerging in the process, forcing software teams to adapt to these changes.  These new roles have resulted in a gap between SE practices, AI communities, and data scientists.  Nalchigar et al.~\cite{nalchigar2021modeling} explains that the gap between stakeholders and data scientists can result in mismatches between business needs and data needs.   On the other hand, Challa et al.~\cite{challa2020faulty} reports that the RE community is not equipped to handle the vast amounts of data needed in building AI systems.  In \cite{altarturi2017requirement}, the authors emphasize the importance of including the data scientist in the process of defining and eliciting requirements, especially requirements related to data extraction.  For AI systems that are data-centered, it is crucial to have a data scientist work closely with the requirements engineers to elicit and identify relevant data for the project.  Therefore, there should be some type of communication between the requirements engineer and the data scientist, especially during the early phases of RE.

\textbf{The Emergence of new Requirements and Techniques:}  With the emergence of new requirements for AI systems such as data, ethics, trust and transparency, new challenges are born for RE.  Bosch et al.~\cite{bosch2018takes} emphasized the need to adapt and complement old practices and techniques with new ones rather than replace the old practices entirely.   Some studies noted that research on RE4AI is not applied to practice, and findings are not being used or addressed by other researchers.  For example, Shin et al.~\cite{shin2019data} identified some data requirements that should be used in AI systems for energy consumption.  However, they found that similar projects were not practicing the use of such requirements.  

The same goes for ethical requirements.  Aydemir and Dalpiaz ~\cite{aydemir2018roadmap} argued that ethics is usually overlooked, and with the change in today’s software systems and the introduction of AI, ethical requirements need to adapt to these changes.  The authors argue that ethics is widely discussed for AI systems but neglected during the building process.  Kuwajima et al.~\cite{kuwajima2019adapting} noted that most software standards such as ISO/IEC 25000 series did not apply to ML systems and has no support for ethical requirements. 

\subsubsection{Issues with data requirements }  One of the significant issues with data requirements is the expense that comes with data-generation \cite{shin2019data}.  Then there is the availability, quality \cite{weihrauch2018conceptual}, training and testing of data \cite{nakamichi2020requirements}.  Requirements need to make sure the quality of data is appropriate, whether the data is available, how to test it, and which data to select for training.  Altarturi et al.~\cite{altarturi2017requirement} explain that RE methods focus on requirements that are user-centric and do not give enough attention to data requirements.  In \cite{ries2021mde} the author argues that Deep Learning (DL) models usually rely heavily on datasets and are not addressed in RE.  The study tries to identify requirements needed for the structure of the datasets and involves finding datatypes, attributes, and properties that need to be elicited during RE. 

The emergence of data requirements has posed new issues for RE.  Sandkuhl~\cite{sandkuhl2019putting} found that data needed for an AI project was easily accessible for many companies.  However, the available data lacked the structure and rules that were necessary to implement and train an AI system.  The study then listed numerous AI requirements to consider.  These requirements included data quality, structure, and format.  For AI-based software that uses data as a primary driver to build the system, it is essential to set rules and carefully select requirements for data selection and management.


\subsubsection{Non-functional Requirements}  

Our understanding of non-functional requirements (NFR) has changed with the addition of AI components when building software systems, and the traditional approaches for managing NFR's require new methods and solutions. These methods need to evolve and re-evaluate how NFR fit into RE4AI.  For example, some NFR's, such as compatibility and modularity, are not as important in ML systems as they were in traditional software systems.  In contrast, other overlooked requirements such as fairness and transparency hold more value \cite{horkoff2019nonFunctional, habibullah2021non}.    There also appears to be less research on modeling non-functional requirements, and research tends to focus mainly on functional requirements~\cite{gruber2017integrated}.

\subsubsection{Requirements for explainable AI}

Providing requirements for explainable AI has created new issues in RE4AI.  For example, in the case of complex systems such as autonomous cars, it can get challenging to specify which parts of the system need to be explained or how to explain them \cite{schwammberger2021quest}.  The author in \cite{hall2019systematic} points out that explainability requirements could collide with risk factors of an AI systems such as in safety-critical systems.  Also, explainability requirements can conflict with others, such as security, cost, and precision\cite{schwammberger2021quest}.  Having an AI system that is more explainable might be more expensive to build.  In such a case, when would it be worth the expense to have a more explainable system?  Kohl et al.~\cite{kohl2019explainability} proposed to use the Softgoal Independency Graph (SIG) to model explainability along with the other NFR to minimize conflicts.


\section{Threats to Validity}~\label{sec:Threats}

In all phases of our SLR, we considered and attempted to mitigate potential threats to validity, common in SLRs for software engineering~\cite{zhou2016map}. 

\textbf{Internal Validity:} To reduce any selection bias while conducting our search, we had an initial protocol to identify a comprehensive set of keywords that were considered relevant in RE4AI literature and tested them out first.  We chose six databases for our search to broaden our results and ensure most studies were included and only selected peer-reviewed papers.  Pilot tests were performed and checked by second and third authors to validate the results.  Snowballing was also performed to capture papers that might have been missed in the initial search results.

From the initial selection criteria, we manually read all the titles and abstracts to filter them further.  The first author did the first round of selection and then verified the results in consultation with the second, third and fourth authors to reach the consensus for the final results.   Once the final list was established, we performed a more detailed scan of the entire document for the resulting papers.   The second filtration process involved a more comprehensive quality assessment test to only include primary studies that passed a specific grade or were in scope.  These studies were selected based on the criteria we set, and for some papers, we were not sure if they fit our criteria.  In such situations, we discussed the paper's selection in several meetings among the authors to ensure the selected article's focus was on RE4AI and to reduce researchers' bias.  A similar process was followed for data extraction and analysis to reach a consensus on the results that were used to answer the research questions.  Another concern for internal validity is publication bias.  To mitigate the publication bias concerns, we checked each primary study as part of the quality assessment criteria to see if there are any reports on any issues and what reliability measures they have performed.  We did observe that out of the 30 empirical studies,  9 did not report negative results or address validity issues.  However, all the selected studies passed the overall quality criteria.

\textbf{Construct Validity:} The major concern for construct validity is based on the appropriateness of the papers selected.  We note that we selected papers that were relevant for our RQs, or if the application was entirely focused on RE4AI.  We further had clearly defined inclusion and exclusion criteria, developed over extensive discussion sessions and relevant to our RQs.  The time frame of our SLR was between 2010 and mid-2021, so any study outside our time frame would not have shown in our results.   

\textbf{Conclusion Validity:} One of the major threats to conclusion validity in SLRs is the bias in data extraction. In our study, during data extraction, codes and themes were based on our RQs, so the data selected was entirely focused on answering our questions.  In order to reduce conclusion validity, we used Nvivo to extract the data by using thematic analysis.  This allowed us to group all results based on the pre-defined codes and themes.  We also found some emerging codes in the process of data extraction, so they were added as needed.  The coding process was done by the first author, so to reduce bias we conducted regular meetings between the first, second and third authors to discuss the data extraction and analysis process to agree on which data should be used and how it would be displayed. 

\section{Discussion and Research Roadmap}~\label{sec:Discussion}

We found from our analysis of the literature that using existing RE techniques for AI-based software development could be challenging. This is primarily due to the different nature of traditional and AI-based software development processes, which has led to new gaps in RE processes for engineering AI software. In this section, we present the gaps that we identified in the selected RE4AI studies, and the recommendations based on these gaps as suggestions for practitioners and future research. Table~\ref{table:Roadmap} presents an overview of these recommendations and maps them to the issues presented in the literature.  Below we elaborate on seven recommendations, presented in the table.

\subsection{Identify the Need and Feasibility of AI} 

The first key issue identified in our analysis of the literature was that many organizations are adopting AI just because ``everyone else was'' and without understanding what the AI component could provide or how much it could solve. Therefore, before deciding to use AI as a software solution, its need should be established first.  These needs include the type of AI solution, data availability and having the resources and expertise to build and manage such systems \cite{GooglePair2019}.  Is AI required to provide predictions, personalize, or make recommendations?  Or is there a need for speech and language understanding, image recognition, or fraud detection? In \cite{schoonderwoerd2021human} the author stresses that before starting an AI project, we need to establish who the users are, what tasks they will perform, and what benefits they would gain from using the system. Other aspects of the AI system should be discussed early on such as understanding what the systems capabilities and limitations are. 

\begin{tcolorbox}[arc=0mm,width=\columnwidth,
                  top=1mm,left=1mm,  right=1mm, bottom=1mm,
                  boxrule=1pt]
We propose that \textit{practitioners} maintain a checkpoint to note all required elements needed to create an AI software system.  The checkpoint should include the problem it is going to solve, why is it required, and how will it be used. And finding out if the organization has all the resources needed to build the AI product.
\end{tcolorbox}

\begin{table}[]
\caption{Mapping of the recommendations to the issues presented in literature} 
\label{table:Roadmap}
\centering
\footnotesize
\begin{tabular}{p{5cm} p{5cm} p{5cm}}
\hline
\textbf{Recommendation} & \textbf{Issue presented in the mapping study} & \textbf{Suggestions for further research} \\ \hline

Identifying the need for AI &  Overconfidence in Using AI & Maintain a check point to list all required elements needed to create an AI-based software system. \\ \hline

Specifying requirements for AI systems & Defining requirements - Non-functional requirements - Emergence of new requirements - Issues with data requirements & Construct a reference map that would capture the key components and attributes needed when specifying AI system requirements  \\ \hline

Using existing RE tools to build AI software & Nature of Machine Learning systems vs. traditional approach of RE - The emergence of new Requirements & Create a taxonomy that gathers all the new techniques and methods for creating AI software \\ \hline

How do we decide on what modeling language to use? & Modelling non-functional requirements & Extend a modeling language to support RE4AI  \\ \hline

How do we bridge the gap between requirement engineers, data scientists, and ML specialist & Existing gap between data scientists and software engineers - Requirements Engineer Responsibilities & Create a platform to share and visually present requirements \\ \hline

How do we address issues such as trade-offs or provide techniques to new RE methods & Calculating the trade-off & Trade-offs should be displayed along with requirements when modeling \\ \hline

Empirical evaluations on Ethics, Explainability and Trust  & The Emergence of new Requirements and Techniques - Requirements for explainable AI & Conduct more empirical evaluations in future research  \\ \hline 
Limited studies on RE for Human-centered AI  &  & Use industrial guidelines to create a framework to include RE for human-centered AI.  \\ \hline 
\end{tabular}
\end{table}

\subsection{Requirements Specifications for AI Systems} 

In \cite{martinez2021software} the authors found that requirements can be hard to specify for AI-based software due to the issues related to measuring and defining requirements for non-deterministic systems.  Also, the emergence of new requirements such as data, ethics and explainability has posed issues to requirements specifications. How do we specify ethical requirements or explain decisions made by a self-driving car?  For example, a vehicle might suddenly brake in front of a bus instead of changing lanes.  This decision would be because it weighs between injuring four people crossing the road versus the one person driving the car getting hurt.  Would the driver make the same decision or disagree with the ethical choices the car company has made on their behalf \cite{cysneiros2018software}?  What requirements do we need to provide in such cases?

Specifying requirements for data has also proven to be a challenge.  First we need to identify the type of AI-based software solution used for the data collected.  Data requirements differ depending on the AI component used.  For example, in supervised machine learning, it is important to have data with learning features and proper labels, whereas an unsupervised model trains on unlabeled data. Specifying key characteristics of data should be set early on to avoid discrimination and biases.  Discrimination can happen when data collection does not include minority groups.  Human-labelled data can also produce biases \cite{vogelsang2019requirements} and using existing data can make it difficult to explain why a given prediction is provided \cite{kohl2019explainability}.  In other situations, such as in safety-critical situations, data needs to be carefully collected and selected. For example, for a medical-surgical robotic application, a requirement would be to find the accurate exertion force for the needle to enter soft tissue.  Data needs to be collected by carefully setting up several experiments~\cite{bonfe2012towards}.  We also found limited studies in our results that focused on identifying and specifying requirements for AI systems.   

\begin{tcolorbox}[arc=0mm,width=\columnwidth,
                  top=1mm,left=1mm,  right=1mm, bottom=1mm,
                  boxrule=1pt]
We recommend that \textit{researchers} should construct a reference map to document requirements for AI systems.  The reference map should capture key components and attributes needed when specifying AI system requirements. We propose using research from the industry and literature to plan and list all possible requirements for AI systems that can be used as a guide to map any emerging requirements into the reference map.  The map may be broken into separate sub-maps to ensure all elements are captured, or can be used for specific AI-based systems, e.g., agent-based systems or NLP solutions.
\end{tcolorbox}




\subsection{Using Existing RE Methods to build AI Software} 

\subsubsection{Existing RE Tools used in RE4AI}

Engineering AI-based software has  had an impact on the way existing tools and techniques are used in RE. Some of these existing tools are not be applicable in RE4AI and the way requirements are elicited and obtained can change when dealing with AI software. In some cases, traditional approaches that require human intervention in gathering data such as interviews and questionnaires are now being replaced by new forms of data collection such as online forms, social media \cite{kostova2020interplay}, sensors~\cite{wang2010networked}, and immersive techniques~\cite{wang2021virtual}. Data collected from such sources requires new RE techniques to elicit and manage them.  There has been an increase in building tools to manage AI software such as the ``AI Playbook''~\cite{hong2021planning} to identify and reduce failures in AI through early AI prototyping, and HINT ``Human-AI INtegration Testing'' \cite{chen2022hint} that automates tests for user interactions with AI systems. Despite these advancements, these tools are still in their early stages; they need further testing and need to be validated in RE context~\cite{martinez2021software}.

\begin{tcolorbox}[arc=0mm,width=\columnwidth,
                  top=1mm,left=1mm,  right=1mm, bottom=1mm,
                  boxrule=1pt]
We recommend that \textit{researchers} should identify the available tools used in building AI software, and determine which ones could be used directly or can be customized for RE.  We also need to build new tools or extend existing ones to support in eliciting, modeling, specifying, and managing requirements for building AI solutions. 
\end{tcolorbox}

\subsubsection{Modeling Languages used in RE4AI}  

Although around 60\% of the selected primary studies demonstrated the use of a modeling language to support requirements, most of the available modeling languages still lacked proper support for RE4AI.   Languages such as UML and GORE were more popular, but had their limitations.  GORE is difficult for non-software engineers to use, considering that team structure in building AI software involved other roles such as data scientists, project managers and ML engineers. Therefore, it becomes more challenging for non-software engineers to learn and use GORE.  In contrast, UML was chosen for its ease of use.  However, UML is not as flexible when modeling non-functional requirements and business rules, which are a core part of RE4AI. 
For example, \cite{aydemir2018roadmap} proposed developing a modeling language that would capture ethical requirements, and \cite{horkoff2019nonFunctional} suggested one to capture NFRs, specific to ML systems. 

\begin{tcolorbox}[arc=0mm,width=\columnwidth,
                  top=1mm,left=1mm,  right=1mm, bottom=1mm,
                  boxrule=1pt]
Future research needs to be done by \textit{researchers} to extend or augment existing modeling languages to aid in capturing and presenting AI requirements, and cutomization is required for different types of AI systems, such as ML systems, agent-based systems and robotic systems.
\end{tcolorbox}
 
\subsection{The Communication Gap Between Requirements Engineers, Data Scientists, and AI Stakeholders} 

As new development team roles and responsibilities emerge when engineering AI software, we found that in order to build proper AI-based software these roles have to communicate information correctly among them.  Currently, there is a lack of communication and integration between roles such as, data scientists, AI stakeholders (e.g., ML engineers and conversational NLP developers) and software engineers~\cite{lwakatare2019taxonomy, vogelsang2019requirements, nalchigar2021modeling}.    For instance, in~\cite{vogelsang2019requirements} the authors observed that in current ML systems, data scientists were responsible for writing high-level requirements, resulting in practices that focus on data and model testing rather than understanding the business domain and stakeholders' needs.  The authors emphasized that the job of the requirements engineer should be eliciting data requirements, while maintaining data provenance, avoiding biases and validating requirements as data might change over time.  At the same time, requirements engineers need to work closely with data scientists as they do not have the expert knowledge to handle and maintain large amounts of data~\cite{challa2020faulty}.  Based on our investigation, data scientists, AI stakeholders and requirements engineers need to improve their knowledge and understanding of the issues arising from blending AI into software systems, and there should be some form of continuous communication among them to set and manage requirements~\cite{amershi2019software,martinez2021software}.  

\begin{tcolorbox}[arc=0mm,width=\columnwidth,
                  top=1mm,left=1mm,  right=1mm, bottom=1mm,
                  boxrule=1pt]
We propose that \textit{practitioners} should create or utilize existing platforms to share and visually present requirements. These platforms should allow all sides of the building team to collaborate, and share ideas and tools in an environment that could enable aspects of RE and AI to be linked and traced. The platform should also allow to capture and present the requirements from different stakeholders' perspective.
\end{tcolorbox}


\subsection{Addressing Issues Related to Calculating the Trade-off}

Trade-offs should be calculated in order to prioritize the importance of different AI-based software requirements.  Google provides a set of guidelines for creating human-centered AI and indicated the importance of weighing different  trade-offs, especially in the case of predictive AI systems. For instance, an incorrect prediction in diagnosing a cancer patient would have more significant stakes than providing a movie recommendation that the user does not like.   When calculating trade-offs, what other requirements can make up for the lost cause?  For example, the study in \cite{shin2019data} experimented with ML algorithms to find which one could produce better results with lower costs.  They found that specific ML algorithms performed better than others. So in such cases, the trade-off could be replaced with other measures that can make up for the loss.  Another study explained that some algorithms provide more reliable predictions but are not easily explained.   Whereas others can better explain why a prediction is delivered, but predictions are less in confidence\cite{krause2016interacting}.   So how do we decide on which algorithm to choose?  In what situations do we prefer to use explainable algorithms vs higher confidence. 

Google PAIR pointed out the importance of calculating the trade-off in the reward function, and to evaluate and weigh the risks of choosing an appropriate reward function that would suit the user's needs. For example, in an ML model that uses classification such as a notification system in an autonomous car, a false negative would \textit{not} notify a sleeping driver in case of an emergency (weightage to precision), which could lead to deadly consequences.  While having too many notifications that are false positives (weightage to recall) can lead the driver to ignore them \cite{dimatteo2020requirements}.  When building AI software, we should always calculate the trade-off between precision and recall, or any other relevant metrics for a given AI solution. When can we choose precision over recall or vice versa?  For example, Dimatteo et al.~\cite{dimatteo2020requirements} explained that in the notification system finding alternative human-centered ways to engage the driver, using recall might be a more feasible and safer choice. Furthermore, Berry \cite{berry2022requirements} reports that the choice of the reward function should also reflect on how the human would perform the task manually and how much impact and value would the chosen reward function provide in return.

Another common trade-off is that some requirements might contradict others (either positively or negatively).  For example, ethics, trust, and transparency can conflict with privacy, safety, and security \cite{cysneiros2020non}.  How do we calculate the trade-off between transparency and privacy?  One might consider the trade-off depending on the application domain.  So a transparent recommendation system might not affect privacy. However, a transparent medical application might reveal some private information that might go against the rules and regulations of the organization.  The trade-offs should be calculated carefully in order to prioritize the importance of requirements. When do we decide on what to choose?  Which model fits best to the organization's needs?  How is the trade-off calculated?  These are all questions that should be addressed during the initial phases.  

\begin{tcolorbox}[arc=0mm,width=\columnwidth,
                  top=1mm,left=1mm,  right=1mm, bottom=1mm,
                  boxrule=1pt]
We recommend that \textit{practitioners} list all trade-offs and the corresponding choices with the rationale for the decisions made.  These trade-offs  should be made explicit along with the requirements.  Also we recommend that \textit{researchers} include trade-offs in modeling languages for RE4AI.
\end{tcolorbox}

\subsection{Lack of empirical evaluations on Ethics, Explainability, and Trust} 

In 2018, the European Commission formed a set of ethical guidelines for trustworthy AI~\cite{HLEG2019ethics}.  The guidelines emphasized that to create trustworthy AI, the outcome should be lawful, ethical, and robust.  Lawful means that developers should comply with all legal regulations.  For example, it should abide by the European General Data Protection Regulation (GDPR) rules and regulations.  Ethical, concluded that the system should have respect for humans, prevent harm, be fair and explicable.  Finally, robust meant that delivered systems should be safe, secure, and reliable.  With the introduction of these guidelines, more research is growing towards creating trustworthy and responsible AI software.  However, certain concepts, such as ethics are not clearly defined and are thus difficult to apply~\cite{cysneiros2020non}. There is a scope for more and in-depth investigation in this area.

Explainable AI systems can help build up users' trust and strengthen their ability to form a more accurate mental model of the product \cite{de2017people}.  However, they are still not applied to AI development as they should.  Amershi et al~\cite{amershi2019guidelines}. identified eighteen guidelines for human-centered AI interaction by examining research from over 20 years of human-centered interaction with AI systems.  The study involved over 150 AI design recommendations collected from research and industrial sources.  The study demonstrated that most violations made in line with the guidelines in current AI systems were linked to explainability.  

People usually expect intelligent systems to provide an explanation similar to how a person behaves and respond \cite{de2017people}.  Miller argues that most developers provide explanations based on their own intuition or understanding of what a good explanation would be, rather than explaining from a social science perspective \cite{miller2019explanation}. Explainable AI could benefit from implementing models that have been used to generate explanations in social and behavioural science research.  However, there is not much research that uses any of these models \cite{miller2017explainable}.   Existing models such as Prospector, ``an interactive visual analytic system" provides explanations to black-boxed machine learning predictions that are hard to explain \cite{krause2016interacting}.  Guidotti et al. also provides a survey listing the different methods used to explain black-box models \cite{guidotti2018survey}. AI developers should benefit from such models and incorporate them into RE practices.

Explaining programs that provide unpredictable outputs can be a challenging task.  However, when given correctly, explanations to predictions can improve people's choices in decision making when it comes to using and adopting AI software \cite{wang2019designing, amershi2014power}.  Our mapping study found limited empirical work on explainability and none for ethics and trust.  

\begin{tcolorbox}[arc=0mm,width=\columnwidth,
                  top=1mm,left=1mm,  right=1mm, bottom=1mm,
                  boxrule=1pt]
We find the need for future \textit{researchers} to conduct more empirical research on RE for explainable, trustworthy and ethical AI software.
\end{tcolorbox}

\subsection{Lack of studies on RE for Human-centered AI}

Many new software systems are moving towards using various AI-based software components~\cite{amershi2019software}. However, it is still far more common to focus on the technical side than on diverse human-centered aspects, such as different user age, gender, ethnicity, and emotions, which are often ignored in the process \cite{maguire2001methods, schmidt2020interactive, grundy2021impact}. Shneiderman~\cite{shneiderman2021human} explains that the focus on machine autonomy in AI software systems can lead to hidden biases. For example, an ML algorithm used in healthcare favoured white people over people of color after the developers failed to include race as a variable when training the algorithm~\cite{WashingtonPostAIbias}; or NLP chatbots are known to learn social biases because of the underlying generic text they are built on~\cite{lee2019exploring}.
Human-centered aspects in RE (in  general) have gained traction in the past years, such as emotions \cite{miller2012understanding}, gender \cite{burnett2016finding, vorvoreanu2019rom}, age \cite{mcintosh2021evaluating}, and mental and physical challenges \cite{grundy2018supporting}. However, we found limited studies that focused on RE for human-centered AI and most RE4AI studies lacked human-centered aspects when specifying requirements.


We think that identifying the user needs is just the first step in building more human-centered AI-based software systems. We need to identify the human-centered needs for other AI-related tasks such as collecting data, choosing the model, explaining outputs, and providing feedback. For example, how do we make sure that the data collected is inclusive, responsible, and non-biased? We found two studies that focused on human-centered aspects in RE4AI~\cite{bruno2013functional,schoonderwoerd2021human}. The first study~\cite{bruno2013functional} involved analyzing requirements for a social robot and emphasizing on emotion when designing the system.  The robot's purpose was to interact with elderly people via speech recognition in an attempt to slow down the chance of developing dementia. The second study~\cite{schoonderwoerd2021human} presents an RE approach for incorporating context-specific human-centered explanations for AI-generated feedback provided to the end users.  
Despite some preliminary work on the topic, we found limited work overall and in particular empirical evaluations on the topic of human-centered AI.

Organisations such as Microsoft have proposed toolkits to promote ethical and responsible AI software development, e.g., ``The HAX Toolkit Project'' \cite{Microsoft2022HAX}.  Although, the HAX toolkit addresses human-centered aspects related to mitigating failures in human-system-interaction, it is limited to the design phase. We further believe that there is a need to create an RE approach that includes all human-centered aspects when building AI-based software. 

\begin{tcolorbox}[arc=0mm,width=\columnwidth,
                  top=1mm,left=1mm,  right=1mm, bottom=1mm,
                  boxrule=1pt]
We recommend that \textit{researchers} should take into consideration the industrial guidelines for building human-centered AI (in addition to existing human-centered related studies), such as Google Pair~\cite{GooglePair2019}, Apple's guidelines on human-centered AI~\cite{Apple2020} and Microsoft's guidelines for human-centered AI interaction~\cite{amershi2019guidelines, Microsoft2022HAX}.
\end{tcolorbox}

\section{Conclusion}~\label{sec:Conclusion}

AI-based techniques have recently become much more embedded into many software systems and are increasingly used by companies to improve performance and reduce costs.  However, using existing RE techniques for current AI systems are challenging due to the different nature of the development process between traditional software engineering methods and AI-based systems.  Current AI systems show a lack of integration with existing RE tools and methodologies, with limited research on the topic.   In this paper, we have presented a mapping study that has identified  43 studies on RE4AI.  We have analyzed different frameworks, methodologies, tools, modeling techniques, and requirements notations currently proposed for RE4AI.  Our results show that most studies favored UML and GORE to model requirements.  Some favored GORE as it had better support for NFR's and business rules, whereas others favored UML as non-software engineers found them easier to use. We also noticed an increased interest over the past year in using DSM to present requirements.  We found that research presented for the autonomous industry is more established. Whereas work on ethics is more theoretical and has just recently gained attention in the research industry.  Our findings identified many issues and challenges in current RE for AI techniques.  For example, defining requirements, explaining predictions, addressing ethical issues, and issues with data requirements.   Another major issue presented in the literature was the lack of integration between software engineers and data scientists.    We concluded by providing a list of research recommendations for future work.  With the lack of current practices available, there is a need to introduce and research new methodologies alongside integrating existing  RE techniques.  The next step should involve documenting any requirements for AI systems, identifying modeling languages, and creating a platform for requirements engineers and data scientists to collaborate and share their ideas.  In the future, we want to evaluate how these techniques can be adopted between different AI areas and investigate specialisations (customisations or new ideas) that need to be developed to meet the needs in specific AI fields.



\printcredits

\bibliographystyle{elsarticle-num}

\bibliography{refs}

\end{document}